\documentclass[reprint, amsmath, amssymb, aps, prab, floatfix, superscriptaddress]{revtex4-2}
\usepackage{graphicx,pstricks}
\usepackage{graphics}
\usepackage{subfigure}
\usepackage{siunitx}
\usepackage{amsmath}
\usepackage{notoccite}
\usepackage{xcolor}
\DeclareUnicodeCharacter{2212}{-}
\begin{document}
\title{Compensating slice emittance growth in high brightness photoinjectors using sacrificial charge}
\author{W. H. Li}
\thanks{Presently at: Brookhaven National Laboratory, Upton, NY, U.S.A. wli4@bnl.gov}
\affiliation{Cornell Laboratory for Accelerator-Based Sciences and Education, Ithaca, NY, U.S.A}
\author{A. C. Bartnik}
\affiliation{Cornell Laboratory for Accelerator-Based Sciences and Education, Ithaca, NY, U.S.A}
\author{A. Fukasawa}
\affiliation{Department of Physics and Astronomy, University of California, Los Angeles, Los Angeles, CA, U.S.A.}
\author{M. Kaemingk}
\affiliation{Cornell Laboratory for Accelerator-Based Sciences and Education, Ithaca, NY, U.S.A}
\author{G. Lawler}
\affiliation{Department of Physics and Astronomy, University of California, Los Angeles, Los Angeles, CA, U.S.A.}
\author{N. Majernik}
\thanks{Presently at: SLAC National Accelerator Laboratory, Menlo Park, CA, U.S.A}
\affiliation{Department of Physics and Astronomy, University of California, Los Angeles, Los Angeles, CA, U.S.A.}
\author{J. B. Rosenzweig}
\affiliation{Department of Physics and Astronomy, University of California, Los Angeles, Los Angeles, CA, U.S.A.}
\author{J.M. Maxson}
\email{maxson@cornell.edu}
\affiliation{Cornell Laboratory for Accelerator-Based Sciences and Education, Ithaca, NY, U.S.A}

\begin{abstract} 
Achieving maximum electron beam brightness in photoinjectors requires detailed control of the 3D bunch shape and precise tuning of the beam focusing. Even in state-of-the-art designs, slice emittance growth due to nonlinear space charge forces and partial nonlaminarity often remains non-negligible. In this work we introduce a new means to linearize the transverse slice phase space: a sacrificial portion of the bunch's own charge distribution, formed into a wavebroken shock front by highly nonlinear space charge forces within the gun, whose downstream purpose is to dynamically linearize the desired bunch core. We show that linearization of an appropriately prepared bunch can be achieved via strongly nonlaminar focusing of the sacrificial shock front,  while the inner core focuses laminarly. This leads to a natural spatial separation of the two distributions: a dense core surrounded by a diffuse halo of sacrificial charge that can be collimated. Multi-objective genetic algorithm optimizations of the ultra-compact x-ray free electron laser (UCXFEL) injector employ this concept, and we interpret it with an analytic model that agrees well with the simulations. In simulation we demonstrate a final bunch charge of 100 pC, peak current $\sim 30$ A, and a sacrificial charge of 150 pC (250 pC total emitted from cathode) with normalized emittance growth of $<20$ nm-rad due to space charge. This implies a maximum achievable brightness approximately an order of magnitude greater than existing FEL injector designs.

\end{abstract}
\maketitle
\section{Introduction}
High brightness photoinjectors are the electron sources of choice for a wide variety of accelerator applications, ranging from x-ray free electron lasers to MeV-scale electron diffraction and microscopy. Higher brightness performance has the potential to unlock probes with greater spatiotemporal resolution, to enable new modes of operation, or dramatically shrink the cost in size of accelerator components. Critical to the brightness performance in these photoinjectors is the process of emittance compensation \cite{serafiniEnvelopeAnalysisIntense1997, carlstenNewPhotoelectricInjector1989}, in which the phase space angles of each longitudinal slice are asymptotically equalized through a judicious choice of focusing and acceleration. Typical implementations of this technique can result in a recovery of brightness by an order of magnitude or more \cite{carlstenNewPhotoelectricInjector1989, chenAnalysisSliceTransverse2021}.

Provided perfect compensation of the linear space charge forces, emittance growth is then dominated by nonlinear forces which distort the slice phase space. With the exception of uniformly filled ellipsoids and infinitely long uniform cylindrical distributions, all other spatial bunch distributions generate nonlinear forces and are susceptible to slice emittance growth. Ideal distributions with linear space charge forces with zero thermal momentum spread will undergo self-similar evolution in each plane; the space charge forces remain exactly linear in propagation through ideal lenses. In practice, however, the intrinsic momentum distribution from photocathodes is often a 3D Gaussian of finite width in the forward direction \cite{vecchioneLowEmittanceHigh2011, fengThermalLimitIntrinsic2015}. This momentum spread causes the evolution of distributions with initially linear space charge to no longer be self-similar, and thus nonlinear space charge emittance growth is in this sense strictly unavoidable. Its magnitude is then dependent on both the slice phase advance in the photoinjector and the size of the intrinsic momentum spread.  

For typical transverse distributions with density maxima on axis, the slice emittance is not in general a monotonically increasing function of time. Several studies \cite{andersonNonequilibriumTransverseMotion2000, floettmannEmittanceCompensationSplit2017, chenAnalysisSliceTransverse2021} have demonstrated that in general the slice emittance oscillates in both uniform focusing channels and split photoinjectors. This oscillation can be understood intuitively: particles at the slice edge are in general initially less defocused than the particles in the core. And so, relative to the initial distribution, a density enhancement is eventually generated at the beam's edge. This enhancement induces stronger space charge defocusing, which eventually creates a distribution similar to the initial conditions with an underdense and under-defocused region at the edge. The process then repeats. 

For infinitely long beams with weak nonlinearity, this oscillation remains in phase with the slice transverse beam size. Thus an important rule emerges: weakly nonlinear beams have slice emittance minima at beam size minima, which is exactly the condition for linear space charge emittance compensation. Thus, the process of linear emittance compensation naturally compensates for nonlinearities in the beam as well. However, in practice, there are several complications to achieving perfect linear emittance compensation, such as nonlaminar focusing of edge particles and slice size mismatch \cite{floettmannEmittanceCompensationSplit2017}. Furthermore, the slice emittance oscillation is not purely periodic: sufficiently nonlinear forces can cause dephasing between the slice emittance and the projected emittance and the initial distribution does not perfectly reform. In practice, this dephasing can occur in half of a plasma period. Nonlinearities strong enough to exhibit this dephasing phenomenon can arise, for example, from 3D effects in bunches with comparable transverse and longitudinal dimensions in the rest frame or sufficiently large transverse density gradients. In photoinjectors, these nonlinearities can additionally lead to \textit{wavebreaking}, wherein the transverse slice phase space becomes doubled-valued. This wavebreaking has been well studied \cite{liNanometerEmittanceUltralow2012, chenAnalysisSliceTransverse2021, floettmannEmittanceCompensationSplit2017} and forms a density shock at the edge of the beam \cite{zerbeDynamicalBunchingDensity2018}. While the brightness of such a wavebroken beam can be substantially improved by eliminating the wavebroken portion with an aperture \cite{zerbeDynamicalBunchingDensity2018}, the nonlinear forces experienced by the core during the wavebreaking process may remain uncompensated.  

The upper limit of beam brightness is set by the brightness of the electron emission  \cite{bazarovMaximumAchievableBeam2009}, and as this value increases with enhanced electric field at the cathode or by improvements in photoemission momentum spread, new strategies may be required for more precise emittance compensation in future photoinjectors. In this paper, we introduce a new technique to undo slice emittance degradation, where we dynamically utilize a wavebroken shock by focusing it through the core of the bunch before subsequently discarding it with an aperture. This process can be thought of as ``wave-crashing":  the broken piece of the wave is nonlaminarly focused through the origin, while the core focuses laminarly, and the net effect is linearization of the bunch core. We demonstrate this linearization process in simulations of the next-generation ultra-compact x-ray free electron laser (UCXFEL) being developed at UCLA \cite{rosenzweigUltracompactXrayFreeelectron2020}. This beamline utilizes recently developed rf accelerating cavities with gradients of up to 140 MeV/m \cite{tantawiDesignDemonstrationDistributedcoupling2020}, including 125 MeV/m accelerating gradient C-band cryogenic linacs \cite{rosenzweigUltrahighBrightnessElectron2018} and 240 MV/m peak fields in the photoinjector \cite{cahillHighGradientExperiments2018}, allowing the length of the beamline to be reduced by more than a factor of 10, making these compact XFELs much more available at the university scale. In the gun, the high fields allow for the generation of beams with brightnesses potentially 50 times better than the original LCLS design, since the maximum 4-dimensional brightness scales with the first or the three-halves power of accelerating field for the pancake and cigar regimes respectively \cite{bazarovMaximumAchievableBeam2009, filippettoMaximumCurrentDensity2014}.  It is important to note that this linearization process was first discovered in simulation by the genetic optimization of the UCXFEL injector, and it was understood analytically after the fact. 

Beyond the cathode electric field, the maximum brightness is also limited by a photocathode quantity known as the \textit{mean transverse energy} (MTE), which encapsulates the transverse velocity spread of the electrons as they are emitted from the photocathode. The intrinsic 2D emittance of the photocathode can be put in terms of the MTE as
\begin{equation} \label{eq:emittance}
\epsilon_{n,x} = \sigma_x \sqrt{\frac{MTE}{mc^2}},
\end{equation}
where $\sigma_x$ is the spot size of the photoemitting laser and $mc^2$ is the rest energy of the electron.

Typically, FELs such as LCLS have used photocathodes with MTEs on the order of 500 meV \cite{akreCommissioningLinacCoherent2008, shuFIRSTDESIGNSTUDIES2019}. Recent developments in photocathode growth and production have enabled the production of photocathodes with MTEs as low as 5 meV \cite{karkareUltracoldElectronsNearthreshold2020}, corresponding to an initial brightness increase of a factor of 100. 

\begin{figure}
\includegraphics[width=\linewidth]{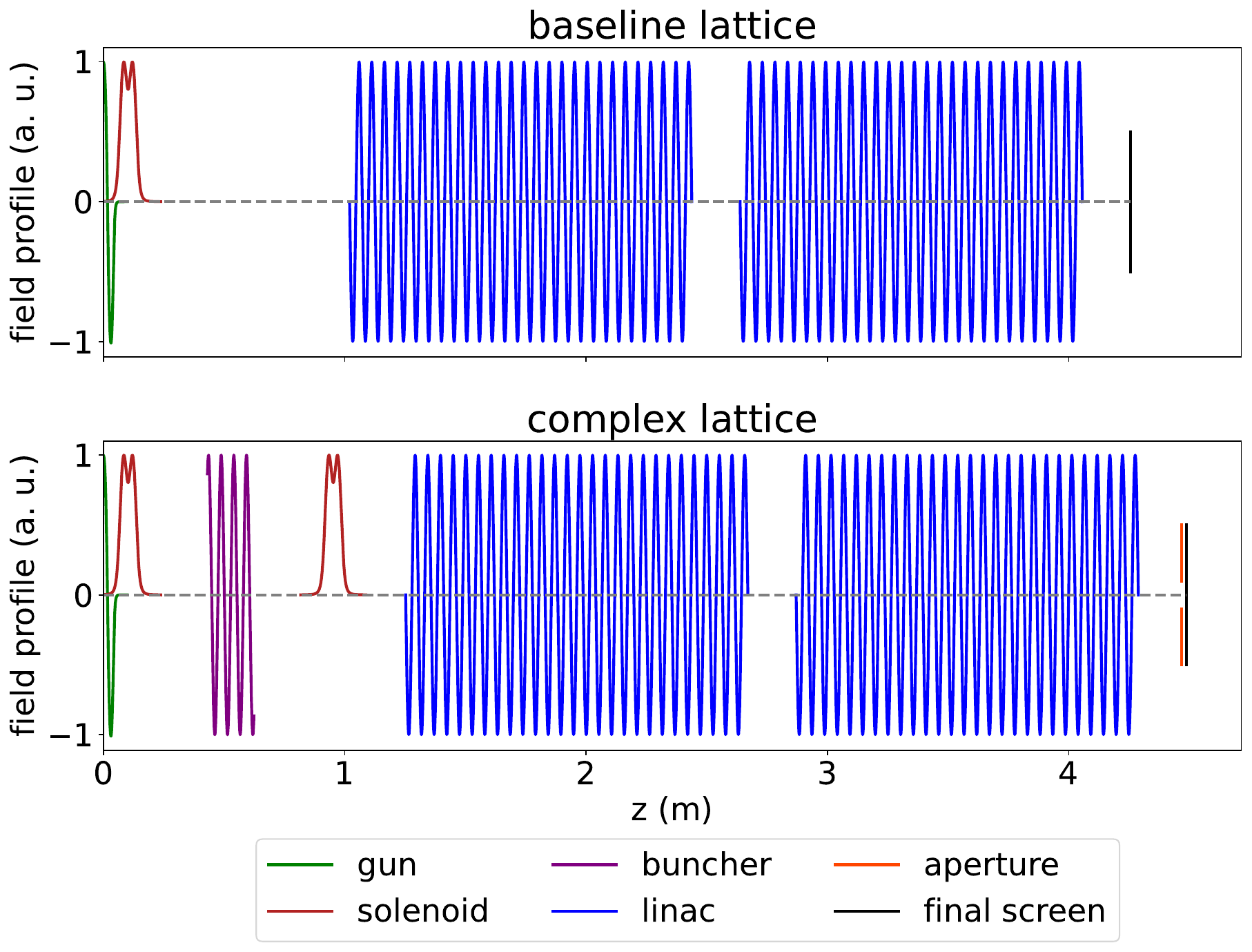}
\caption{Schematic of the model XFEL photoinjector with the field profiles of each element. (a) The baseline lattice. (b) The complex lattice, which can be augmented with an aperture.}
\label{fig:layout}
\end{figure}

Our study begins with a discussion of the performance enhancements achievable when combining the high-gradient accelerating cavities of the UCXFEL beamline and a low MTE photocathode. The associated high initial charge density renders this an ideal environment for studying the dynamics of cold, highly space charge dominated beams. Previously \cite{bartnikOperationalExperienceNanocoulomb2015, gullifordDemonstrationLowEmittance2013, gullifordDemonstrationCathodeEmittance2015}, a lattice with two solenoids and a buncher has been shown to perform well in space charge dominated beamlines. We will examine the performance of such a lattice, comparing its performance with and without a final clipping aperture, and explain the physics behind the qualitatively different dynamics that arise in the presence of the aperture. A simplified lattice with a single solenoid will also be included for reference, and the two lattices are shown in Fig. \ref{fig:layout}. We will refer to the lattice with the single solenoid as ``baseline" and the lattice with two solenoids and a buncher as ``complex".

\section{Emittance preservation}
We use a multi-objective genetic algorithm (MOGA) \cite{gullifordMultiobjectiveOptimizationsNovel2016}, coupled to a particle-in-cell code, General Particle Tracer (GPT) \cite{GPTSite}, to find optimal settings for these three representative high gradient XFEL beamlines and determine the best achievable emittance. In turn, we can use the optimal emittances to determine the degree to which the initial emittance has been recovered.

The final beam energy in all three beamlines is 150 MeV. In the baseline lattice and the complex lattice without the aperture, we will transport 100 pC of charge from the beginning to the end, while in the complex lattice with the aperture, we will start with 250 pC and end with the same 100 pC, clipping down with the final aperture. The cathode MTE is set to 5 meV for all simulations, as this is the lowest that has been experimentally achieved \cite{karkareUltracoldElectronsNearthreshold2020}, as well as being near the fundamental limit imposed by disorder induced heating \cite{maxsonFundamentalPhotoemissionBrightness2013}.

We set the MOGA to optimize both transverse emittance and bunch length. In general, optimizing multiple objectives will generate a curve known as a Pareto front which shows the tradeoff between the objectives. By comparing the Pareto fronts for the baseline lattice and the complex lattice without the aperture, we can see how much of an improvement the extra solenoid and the buncher provides. Subsequently, the effect of the aperture can be identified by looking at the front for the complex lattice with the aperture. 

\begin{figure}
\includegraphics[width=\linewidth]{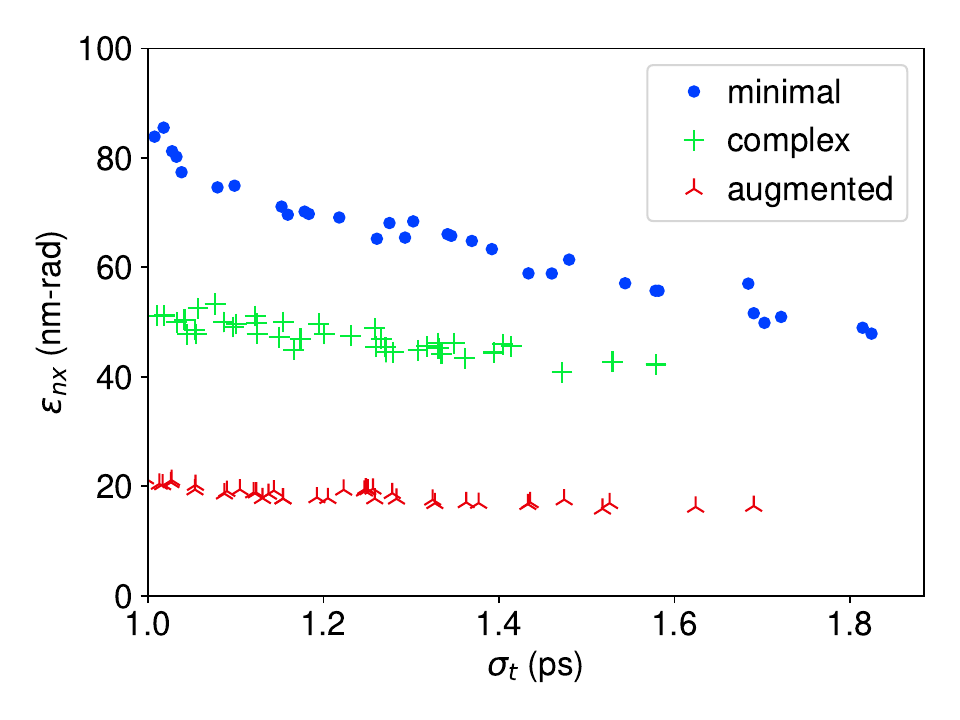}
\caption{Pareto fronts for the three lattice configurations: baseline, complex without the aperture, and complex with the aperture (augmented). The addition of the second solenoid and buncher improves emittance by roughly a factor of 1.4 on average, and the addition of the aperture improves it by more than another factor of 2.}

\label{fig:fronts}
\end{figure}

The fronts are shown in Fig. \ref{fig:fronts}. We see immediately that the complex layout improves the emittances by roughly a factor of 2, with the aperture providing yet another factor of 2. The degree to which the cathode contributes to the final emittance can be quantified by defining an \textit{effective MTE} as the MTE required, given the laser spot size, to generate the observed final emittance assuming perfect emittance preservation\cite{pierceLowIntrinsicEmittance2020}. Thus, the effective MTE is a heuristic to determine the degree to which emittance is preserved during transport and to determine the MTE scale at which the intrinsic emittance becomes dominant. An ideal beamline would have an effective MTE equal to the real MTE of the cathode. Dilution in transport is captured by an increase in the effective MTE.

\begin{figure}
\includegraphics[width=\linewidth]{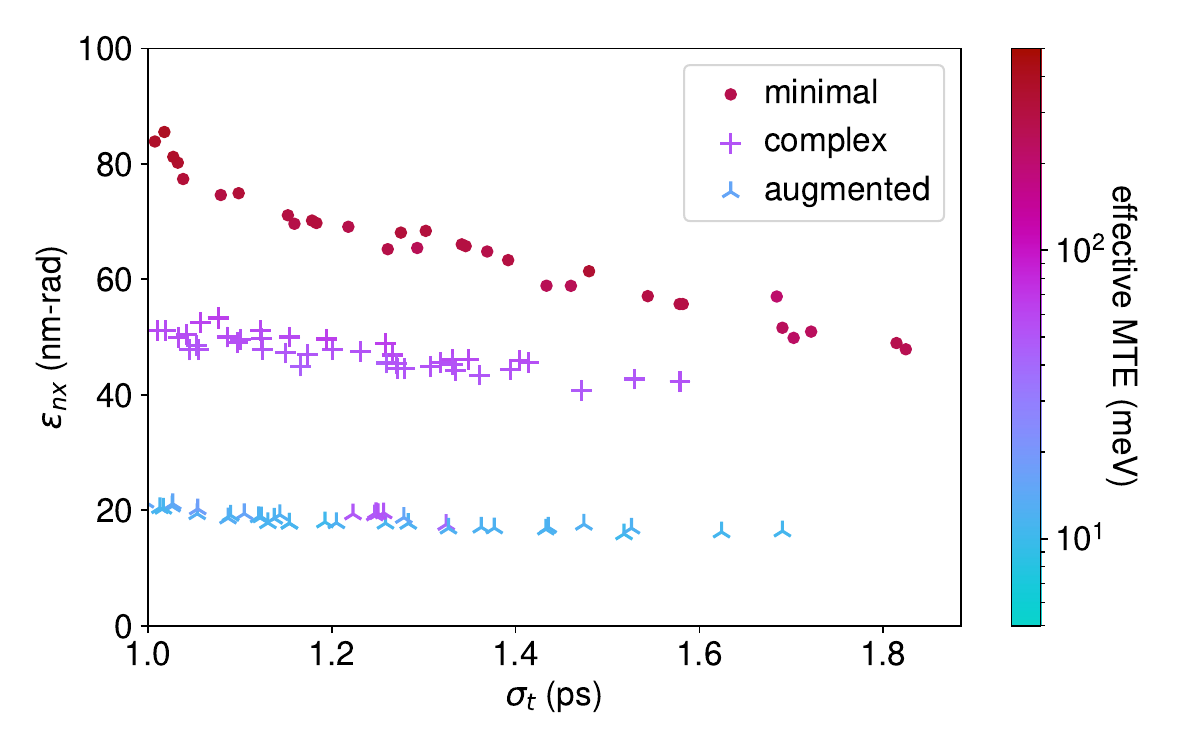}
\caption{The Pareto fronts from Fig. \ref{fig:fronts} colored by their effective MTE (defined in the text). The effective MTE of the baseline lattice is approximately 300 meV, showing a substantial emittance growth. The complex lattice without the aperture achieves an effective MTE on the order of 70 meV on average, while adding the aperture achieves an effective of about 10 meV on average.}

\label{fig:fronts_with_eff_mte}
\end{figure}

The addition of the aperture introduces a small complication to the effective MTE calculation. Directly calculating the effective MTE using the emittance of the core 100 pC but the laser spot size used to emit the full 250 pC would result in an artificial reduction of the effective MTE. Instead, we use the laser spot size which would be required to emit 100 pC with the same charge density as the full 250 pC beam, which can be approximated by scaling the naive effective MTE up by a factor of 2.5, the ratio of the emitted charge to the final charge. The effective MTEs of all three fronts can be seen in Fig. \ref{fig:fronts_with_eff_mte}. The baseline lattice has effective MTEs on the order of 300 meV showing that the MTE is not a significant contributor to the emittance and that the emittance has been substantially degraded during transport. Such a lattice would see little benefit from upgrading from a 100 meV cathode to a 5 meV cathode. The complex lattice without the aperture does better, with an effective MTE of 70 meV. This lattice would be cathode emittance-limited with a 100 meV cathode, but would not be able to utilize a single-digit MTE cathode to its maximum potential. However, the complex lattice with the aperture has an effective MTE of around 10 meV, indicating excellent emittance preservation. We see that to take full advantage of this beamline, a cathode with single-digit meV MTE is necessary.

\begin{figure}
\includegraphics[width=\linewidth]{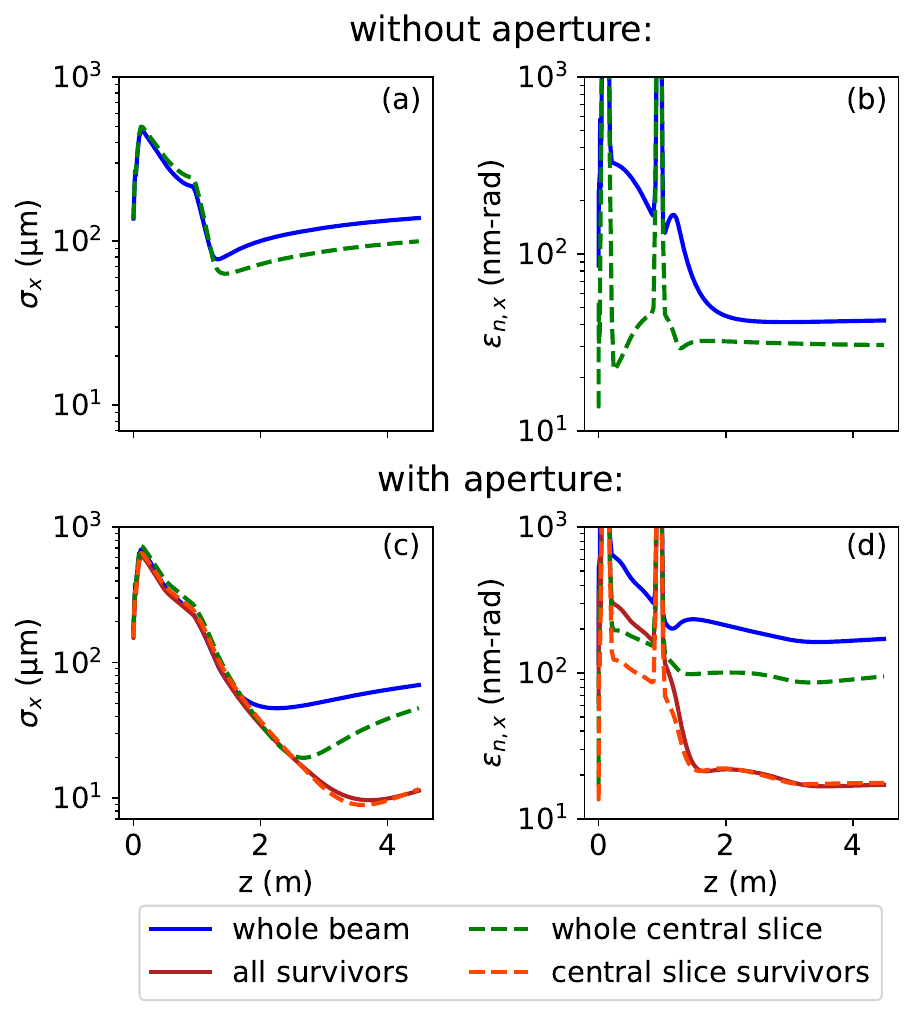}
\caption{Transverse beam size (a,c) and emittance (b,d) evolution of representative individual simulations from the complex lattice without (a,b) and with (c,d) the aperture. The complex lattice without the aperture achieves precise traditional emittance compensation, characterized by the beamsize and slice emittance minima coinciding in longitudinal position. In contrast, with the aperture, the slice emittance increases drastically immediately upon exiting the gun before rapidly falling as the beam enters the linacs. The beam size evolution of the two lattices is also noticeably different.}
\label{fig:emittance_evolution}
\end{figure}

To isolate the effects of the aperture from the rest of the lattice, we will now focus on the two beamlines utilizing the complex lattice. Examining the beam size and emittance evolution of representative individuals in their fronts reveals a stark qualitative contrast in their respective dynamics, shown in Fig. \ref{fig:emittance_evolution}. The complex lattice without the aperture performs precision traditional emittance compensation. Due to residual uncompensated nonlinear space charge effects, the slice emittance never reattains the emittance achieved immediately after the first solenoid. On the other hand, the complex lattice with the aperture allows the slice emittance to rise by nearly an order of magnitude immediately at the first solenoid. The slice emittance then decreases until it is frozen by the linacs. Furthermore, while the beamsize in the complex lattice without aperture reaches its minimum close to the entrance of the first linac, the beamsize in the complex lattice with the aperture reaches a minimum in the middle of the first linac, with the beamsize of the survivors reaching a minimum in the middle of the second linac. Thus, not only is the beam evolution in the two beamlines different, but it also appears that in the beamline with the aperture, there are effectively two separate particle populations.

\begin{figure}
\includegraphics[width=\linewidth]{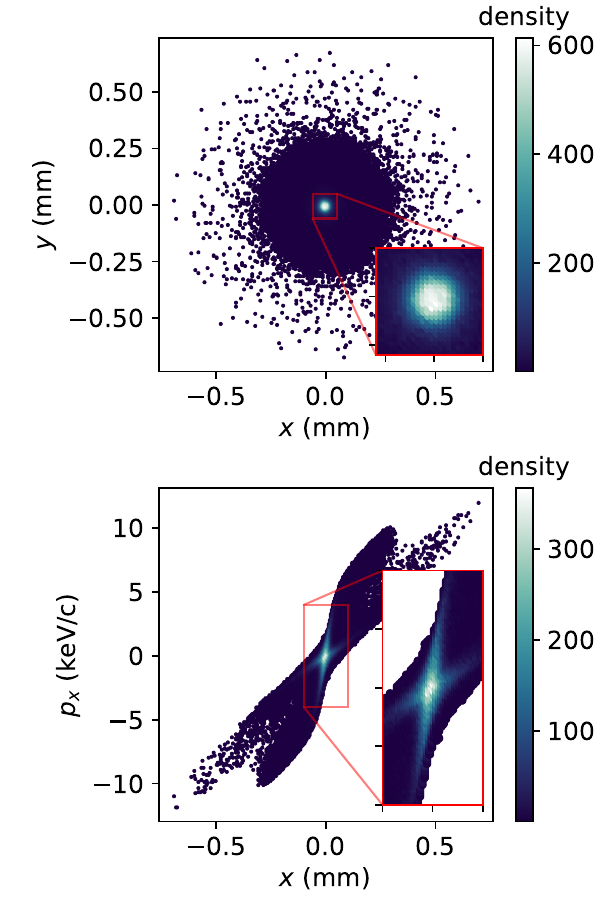}
\caption{The beam profile and $x$ phase space immediately before the aperture. (a) Beam profile with linacs. (b) $x$ phase space with linacs. The beam core has become very dense and is surrounded by a diffuse halo. Insets: zoomed in picture of the core.}
\label{fig:aperture_effect}
\end{figure}

The two populations can be visually identified in Fig. \ref{fig:aperture_effect}, which shows the beam profile and $x$ phase space immediately before the final aperture. A dense core in both real space and phase space has developed, surrounded by a diffuse halo of electrons. The aperture discards this diffuse halo, reducing the emittance from 150 nm-rad to 20 nm-rad while only losing 60\% of the bunch charge. 

\begin{figure}
\includegraphics[width=\linewidth]{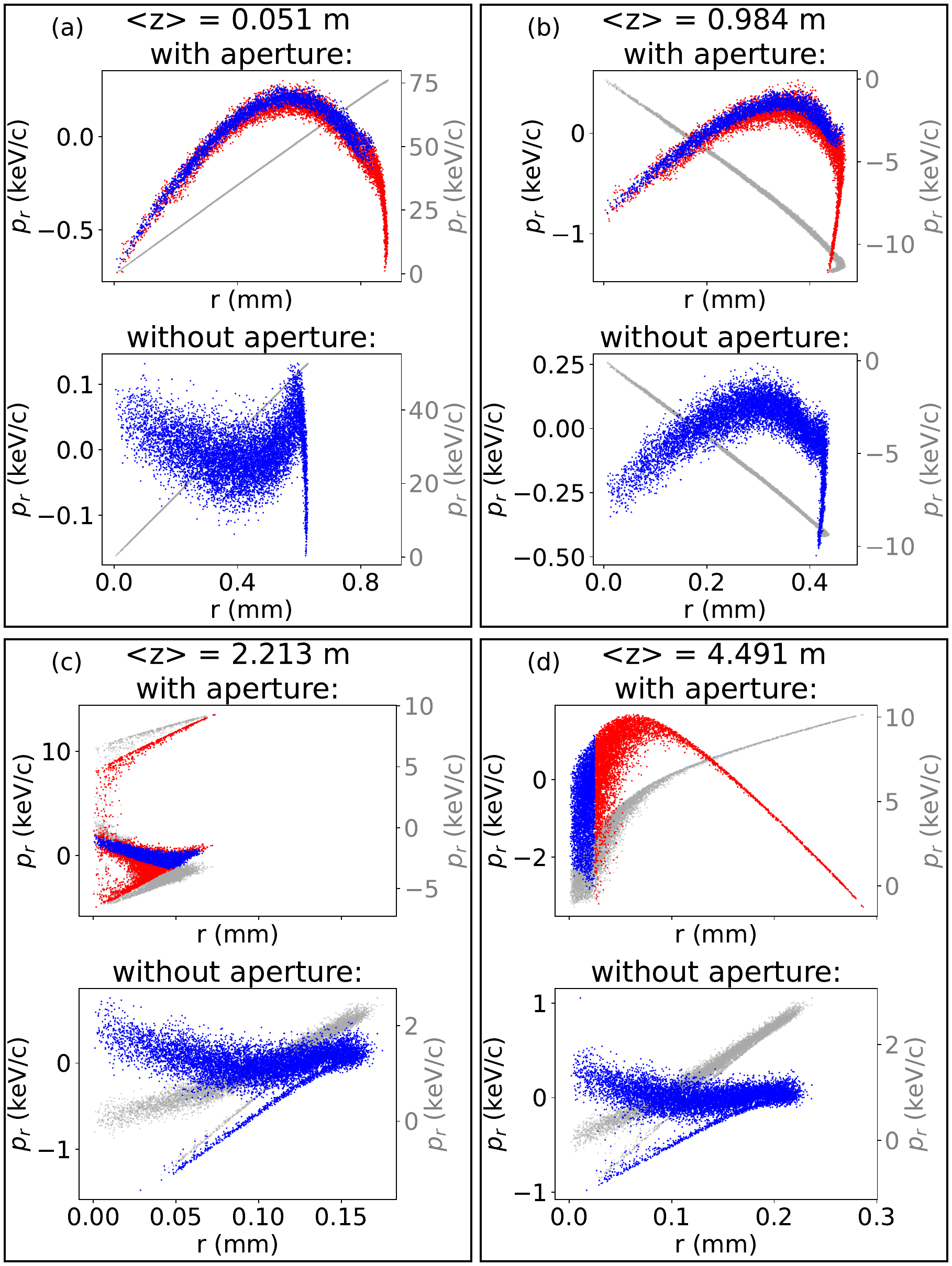}
\caption{The evolution of the $r-p_r$ phase space of the central longitudinal slice of the beam in the complex lattice. Top row of each subfigure: with the aperture. Bottom row of each subfigure: without the aperture. The dead particle (red) and survivor (blue) phase spaces are displayed with the linear $r-p_r$ correlation removed. The original, correlated phase space is shown in light gray. Note that the phase spaces of the two lattices are plotted with the same horizontal axis scale but different vertical scales. (a) Immediately after the exit of the gun. (b) Immediately after the second solenoid. (c) During transport through the linacs. (d) Immediately before the final screen.}
\label{fig:phase_space_evolution}
\end{figure}

To understand the process by which this dense core develops, we look at selected snapshots of the beam's phase space, plotted in $r-p_r$ coordinates, for both beamlines, shown in Fig. \ref{fig:phase_space_evolution}. In all plots, red particles represent particles that are clipped by the aperture, which we call \textit{dead particles} and blue particles represent particles that pass through the aperture, which we call \textit{survivors}. Since the aperture is at the end of the beamline, the coloration reflects knowledge of the eventual trajectories of these particles. As the linear $r-p_r$ correlation is usually large and washes out the details, we plot the phase spaces with this correlation subtracted, with the original correlated phase space being shown in gray for reference. Since the dynamics of the survivors in the central slice and the survivors in the rest of the beam are nearly identical, as can be seen in Fig. \ref{fig:emittance_evolution} (c), we look only at the phase space evolution of the central slice.

Figure \ref{fig:phase_space_evolution} shows the phase space immediately after the gun, where there is already a significant difference between the two lattices. In general, the curvature of the space charge force varies with the beam's aspect ratio \cite{kimRfSpacechargeEffects1989}. Without the aperture, the photoemission process is tuned to balance the varying curvature and generate a phase space without large nonlinearity (and a small wave-broken tail). With the aperture, the space charge force is permitted to generate a large negative concavity in phase space, which eventually results in the shock front forming at the edge of the beam.

The beam then travels through the buncher and the two solenoids, with the phase space at the exit of the second solenoid shown in Fig. \ref{fig:phase_space_evolution} (b). Of particular note here are the much larger radial velocity spread of the outermost particles in the lattice with the aperture, as well as the larger focusing force applied to the beam as a whole. This results in the wave-crashing shown in Fig. \ref{fig:phase_space_evolution} (c), where the fastest falling particles in the shock front have been nonlaminarly focused through the core of the beam. Contrast this to the lattice without the aperture below, where the nonlinear tail has merely been folded under the rest of the beam and no nonlaminarity has occurred.

Finally, Fig. \ref{fig:phase_space_evolution} (d) shows the phase space immediately before the final screen, where the aperture is placed for the lattice that utilizes it. We see again that the lattice with the aperture produces a beam with a very dense spatial core surrounded by a diffuse halo. Meanwhile, the lattice without the aperture has generated a roughly flat phase space, but residual curvature is still visible, as well as the folded nonlinear tail.

\begin{figure}
\includegraphics[width=\linewidth]{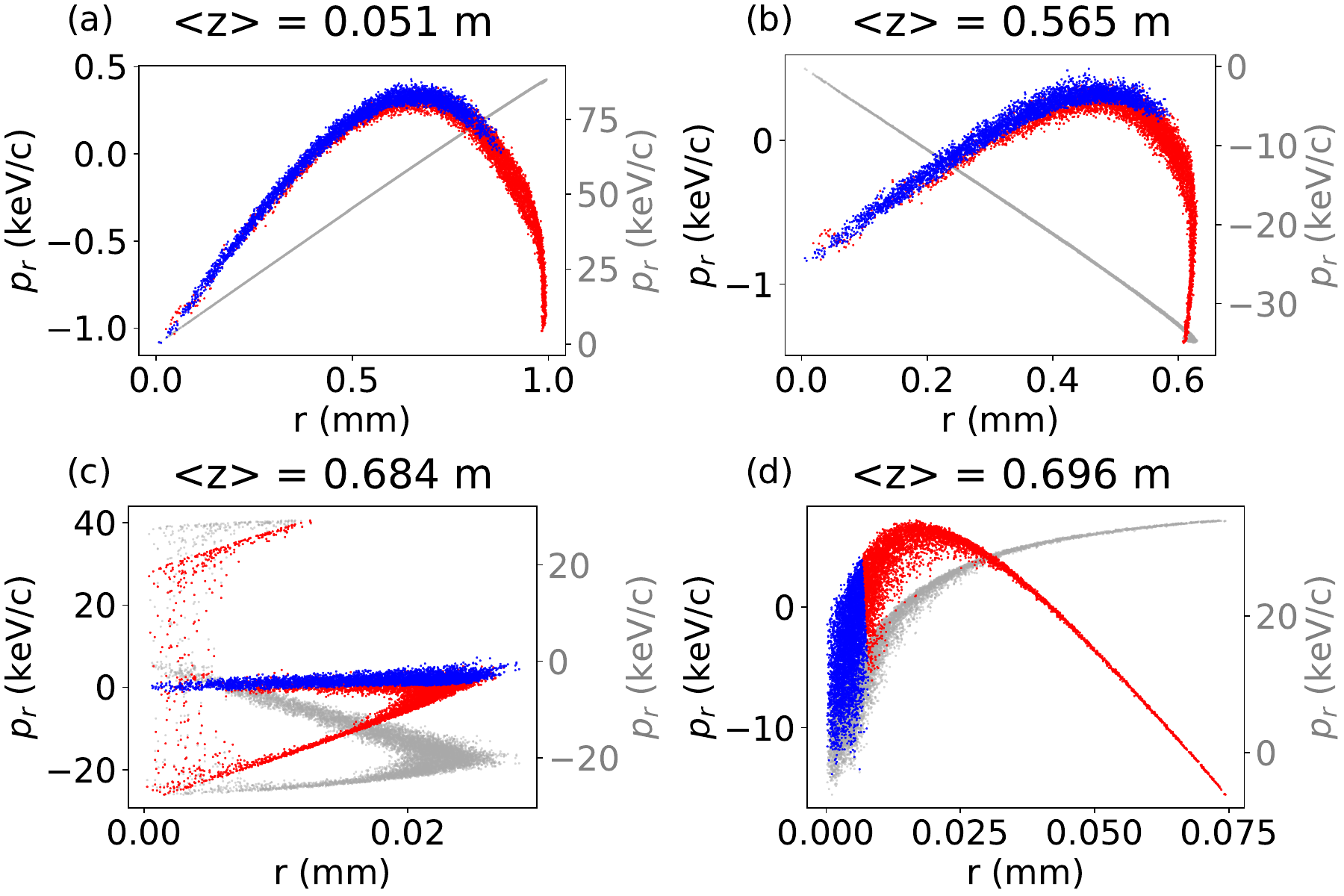}
\caption{The evolution of the $r-p_r$ phase space in an beamline with the aperture reoptimized without the linacs. The coloring has the same meaning as in Fig. \ref{fig:phase_space_evolution}. The similarities in the phase space evolution to the beamline with the linacs are evident. (a) Immediately after the exit of the gun. (b) Immediately after the second solenoid. (c) Transport to the final screen. (d) Final screen.}
\label{fig:no_linac_phase_space_evolution}
\end{figure}

Our goal for the remainder of this paper will be to understand the forces at play during the wave-crashing which linearize the phase space of the core and generate the high core density. To aid in understanding the process by which this dense core develops, we simplify the simulation by removing the two linacs and the buncher and moving the other elements closer to the gun, reoptimizing the lattice and initial electron beam parameters. Figure \ref{fig:no_linac_phase_space_evolution} shows the phase space evolution at approximately equivalent points on the beamline as shown in Fig. \ref{fig:phase_space_evolution}. The basic dynamics are identical, with a large curvature developing inside the gun, a shock front forming at the edge of the beam which is subsequently nonlaminarly focused through the core, and a dense core appearing at the end of the beamline. The following model is based on this linac-less simulation.

\section{Wave-crashing linearization}
This section focuses on understanding the physics that causes these highly nonlinear dynamics to improve, rather than dilute, the emittance of the core. We begin with a simplified analytic model of the space charge forces imparted by the shock front on the beam as it collapses inwards. In the beam's rest frame, it is on the order of 20 times longer than it is wide after exiting the second solenoid, so we treat it as an infinitely long cylinder of charge, surrounded by an infinitely long, infinitely thin, cylindrical shell of charge. At $t=0$, the shell has radius $a_0$, defined to be the radius of the outermost particle in the beam. We will assume the particles that form the shell have velocities uniformly distributed from $v_s$ to $v_f$, all directed radially inwards. We will perform all calculations in the rest frame of the beam. 

The beam is being focused, which we will model as the beam receiving a linear velocity kick $v = -Kr$ at $t=0$, directed radially inwards. From Fig. \ref{fig:phase_space_evolution} (b), we see that all particles in the shell are focusing faster than the particles in the beam, so we have $v_s > Kr$ for all $r$. 

We see from the phase space pictures that the linearization is already completed by the time the fastest particles in the shell reach a focus, and after this point, which we call $t_f$, we will assume that the force from the shell is negligible due to the large decrease in the shell's charge density. We want to find the impulse provided by the shell to a particle with charge $q$ at a start position $\vec{r_0}$, i.e., the force from the shell integrated from $t=0$ to $t=t_f$. As a simplification, we will assume that the velocities of all particles is constant, i.e., the space charge force only provides small perturbations to their trajectory. Thus, the shell will undergo a nonlaminar focus as it collapses past the center and starts expanding outward.

The impulse can be written
\begin{equation}\label{eq:impulse}
\Delta\vec{p} = \int_0^{a_0/v_f} q\vec{E}(t)\,dt,
\end{equation}
where $\vec{E}(t)$ is the field at $\vec{r}(t)$.

To calculate $\vec{E}$, consider the subset of shell particles with average velocity $\vec{v}$ and velocity spread $d\vec{v}$. Since we assume the velocities are uniformly distributed and all point radially inwards, this subset is itself a cylindrical shell with a linear charge density given by
\begin{equation}\label{eq:charge}
\lambda = \frac{\lambda_{r}dv}{v_{f} - v_{s}},
\end{equation}
where we will begin writing only the magnitudes of the velocities and $\lambda_{r}$ is the total linear charge density of the collapsing shell. 

The field from an infinitely long cylinder of surface charge density $\sigma$ and radius $a$ at a point $\vec{r}$ outside the shell is
\begin{equation}\label{eq:field}
\vec{E} = \frac{a\sigma}{\epsilon_0 r}\hat{r}.
\end{equation}

We note that $a\sigma$ represents the linear charge density of the shell and is identical to the previously defined $\lambda$, which is constant with time, so
\begin{equation}\label{eq:field_time}
\vec{E}(t) = \frac{a\sigma}{\epsilon_0 r(t)}\hat{r} = \frac{a\sigma}{\epsilon_0 r_0(1-Kt)}\hat{r}.
\end{equation}

The field inside the cylinder is identically 0, so a subset with a given velocity will only contribute to the field once it has collapsed past $\vec{r}$.

We will define $t_p(v)$ to be the time that a particle with velocity $v$ has a radial position equal to that of the test charge. Thus, it must satisfy the equation:
\begin{equation}\label{eq:t_p}
a_0 - vt_p(v) = r_0(1-Kt_p(v))
\end{equation}
We can solve this to obtain
\begin{equation}
t_p(v) = \frac{a_0-r_0}{v-Kr_0}.
\end{equation}

For $t<t_p(v)$, the field is identically 0, because the subset has a radius greater than the radial position of the test charge. Since the upper limit of the integration is when the fastest shell particles reach the center of the beam and all shell particles have a higher velocity than all beam particles, the subset of shell particles with velocity $v$ has a radius smaller than that the radial position of the test charge for times between $t_p(v)$ and $a_0/v_f$, and its field contribution is non-zero during this time. Thus, the integral in Eq. \ref{eq:impulse} becomes, as a function of the subset velocity $v$,

\begin{equation}\label{eq:impulse_int}
    \Delta\vec{p}(v) = \int_{t_p(v)}^{t_f} \frac{q \lambda}{\epsilon_0 r_0(1-Kt)}\hat{r}dt.
\end{equation}

\begin{figure}
    \includegraphics[width=\linewidth]{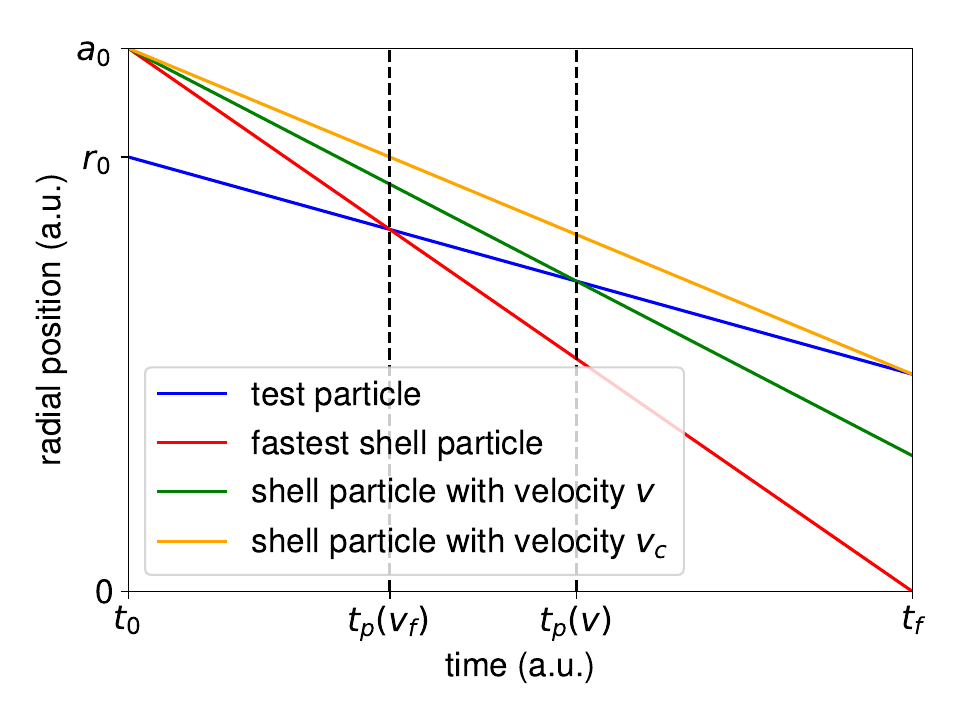}
    \caption{Cartoon of sample particle trajectories and the associated quantities defined in our model.}
    \label{fig:analytic_cartoon}
\end{figure}

We will define the critical velocity $v_c$ to be the velocity for which $t_p(v_c) = t_f$, i.e., a shell particle with velocity $v_c$ attains the same radial position as the test charge at the same time that the fastest shell particle reaches the center of the beam. All particles with velocity lower than $v_c$ will not contribute to the field experienced by the test charge. Note that $v_c$ is a function of the test charge starting position $r_0$. We have
\begin{equation}\label{eq:vc_focus}
v_c(r_0) = v_f - r_0\left(\frac{v_f}{a_0} - K\right).
\end{equation}

Figure \ref{fig:analytic_cartoon} shows a cartoon of particle trajectories to clarify the physical meaning of the quantities we have defined. All shell particles begin with radial position $a_0$, while the test particle begins with radial position $r_0$. $t_f$ is the x-intercept of the trajectory of the fastest shell particle, and $t_p(v)$ is the time when the trajectory of the test particle and the shell particle with velocity $v$ cross. Note that $t_p(v_c)$ is defined to be identical to $t_f$.

We now have two cases: if $v_c < v_s$, then all shell particles will contribute to the impulse on the test charge; if $v_c > v_s$, then some particles have radial position greater than that of the test charge at all times in the integration and do not contribute to the impulse. Therefore, to calculate the total impulse provided to the test charge by the entire shell, we integrate Eq. \ref{eq:impulse_int} setting the maximum of $v_c$ and $v_s$ as the lower limit and $v_f$ as the upper limit:
\begin{equation}
    \Delta \vec{p}(r_0) = \frac{q \lambda_r}{r_0\epsilon_0(v_f - v_s)} \int_{\max(v_c(r_0), v_s)}^{v_f} \int_{t_p(v)}^{t_f} \frac{1}{1-Kt} dt dv.
\end{equation}
These integrals can be carried out analytically, to yield the following piecewise function (note that the subscript on $r$ has been dropped):
\begin{equation}\label{eq:impulse_final}
    \Delta \vec{p}(r) =
    \left\{
        \begin{array}{lr}
              \frac{q \lambda_r}{K r \epsilon_0 (v_f-v_s)} L_s\hat{r}, & r<\frac{a_0 (v_f- v_s)}{v_f - Ka_0}\\
             
             \frac{q \lambda_r}{Kr \text{$\epsilon $0} (v_f-v_s)} L_b\hat{r}, &  r \geq \frac{a_0 (v_f- v_s)}{v_f - Ka_0}
        \end{array}
    \right\},
\end{equation}
where we define
\begin{align}
	L_s \,&= K r \log \left(\frac{a_0(v_f - Kr)}{v_f(a_0-r)}\right) \nonumber\\
	&+ K a_0 \log\left(1-\frac{r}{a_0}\right) + v_f \log \left(\frac{v_f}{v_f - K r}\right)
\end{align}

and
\begin{align}
    L_b \,&= K r \log \left(\frac{v_f - Kr}{v_s - Kr}\right) \nonumber\\
    &+K a_0 \log \left(\frac{v_s-K a_0}{v_f-K a_0}\right)\nonumber\\
    &+v_s \log \left(\frac{(v_f-K a_0)(v_s - K r)}{v_f(v_s - Ka_0)}\right) \nonumber\\
    &+v_f \log \left(\frac{v_f}{v_f-Kr}\right).
\end{align}
Note that $L_s$ corresponds to the case where $v_c(r) > v_s$, while $L_b$ corresponds to the case where $v_c(r) \leq v_s$.

We can now calculate the impulse provided to the beam by the collapsing shell in our simulation. We will consider the timestep shown in Fig. \ref{fig:phase_space_evolution} (b) to be the start of the collapse. Notice that the velocity spread increases by about a factor of 10 between the timesteps shown in (b) and (c). This is due to the slower particles experiencing the space charge field from a larger enclosed charge and being slowed more than the faster particles. For now, we will continue to neglect the space charge force applied to the shell and instead examine the behavior of the beam for a range of velocity spreads. It is worth noting that the large increase in velocity spread indicates that the space charge force is significant and should not be neglected, but, remarkably, the correct behavior arises regardless.

\begin{figure*}
    \centering
    \includegraphics[width=\textwidth]{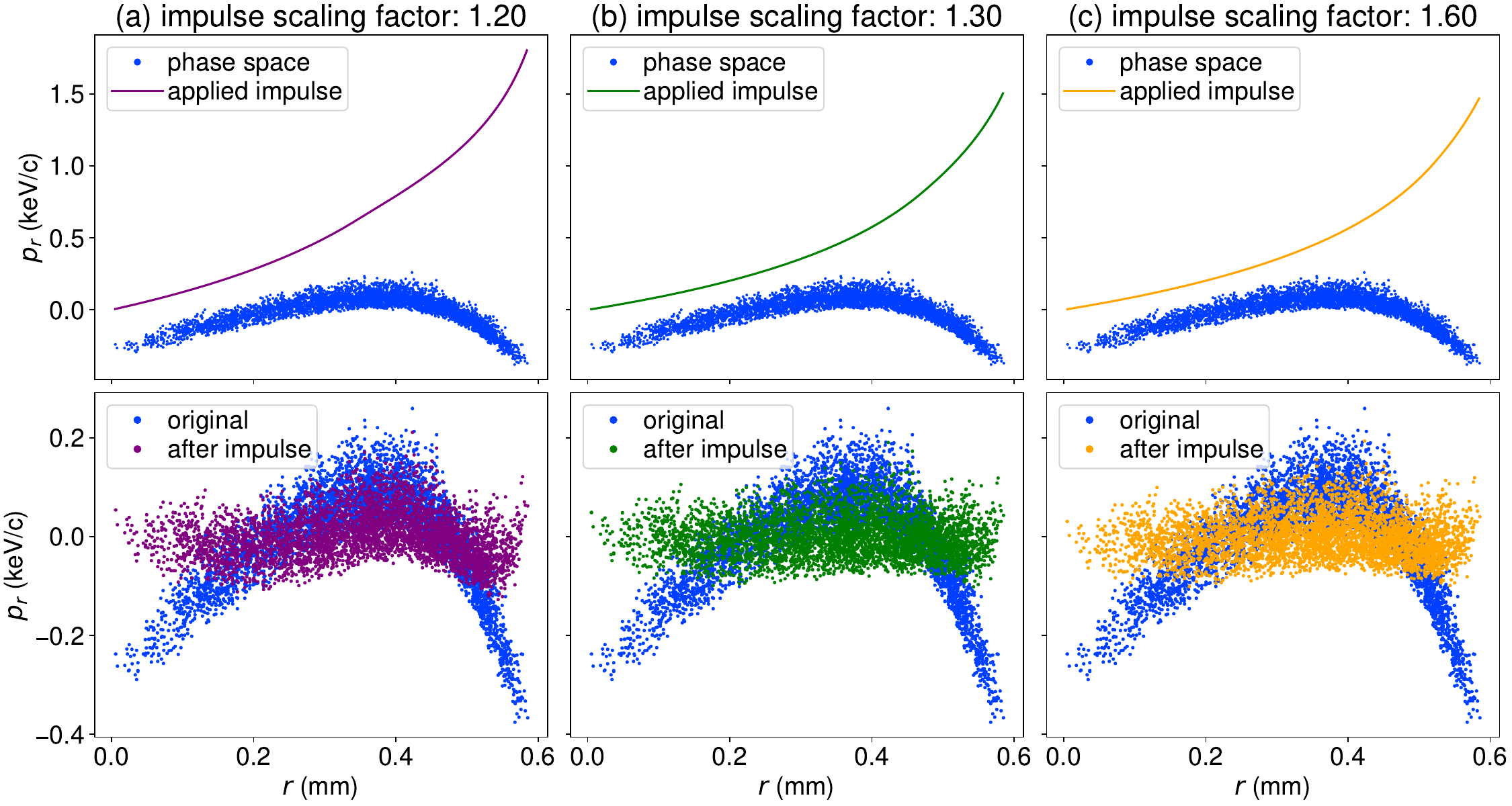}
    \caption{Analytically calculated impulses and final phase spaces. Phase spaces are displayed with linear $r-p_r$ correlations removed. Top row: impulses calculated from Eq. \ref{eq:impulse_final}. Bottom row: final phase spaces. (a) Low velocity spread. (b) Average velocity spread. (c) High velocity spread. The ``impulse scaling factor" is an overall factor multiplying the applied impulse, which can be thought of as representing different optimal shell charge densities in each case.}
    \label{fig:analytic_individuals}
\end{figure*}

The results are shown in Fig. \ref{fig:analytic_individuals} for low velocity spread (half of the average), average velocity spread, and high velocity spread (double the average). The ``impulse scaling factor" is an overall hand-tuned correction factor multiplying the applied impulse. Since the impulse is linear in the shell charge density, this scaling factor can be thought of as representing differences in the optimal shell charge density for each velocity spread. The shape of the impulse is generally the same in all cases and the phase space becomes straightened out. The only change that is needed for a different velocity spread is an overall scaling factor for the impulse, shown in the titles of each plot. The final emittances are 23 nm-rad, 21 nm-rad, and 22 nm-rad for the low, average, and high velocity spreads respectively, down from 53 nm-rad at the exit of the second solenoid. The emittance from simulation is 19 nm-rad, very good agreement considering the many simplifications we have made in the model. The deviance of the scaling factor from unity is largely due to ignoring the effects of space charge on the motion of the shell. As we shall see in the next section, incorporating these effects into the model numerically yields a scaling factor of nearly unity.

\begin{figure}
    \centering
    \includegraphics[width=0.6\linewidth]{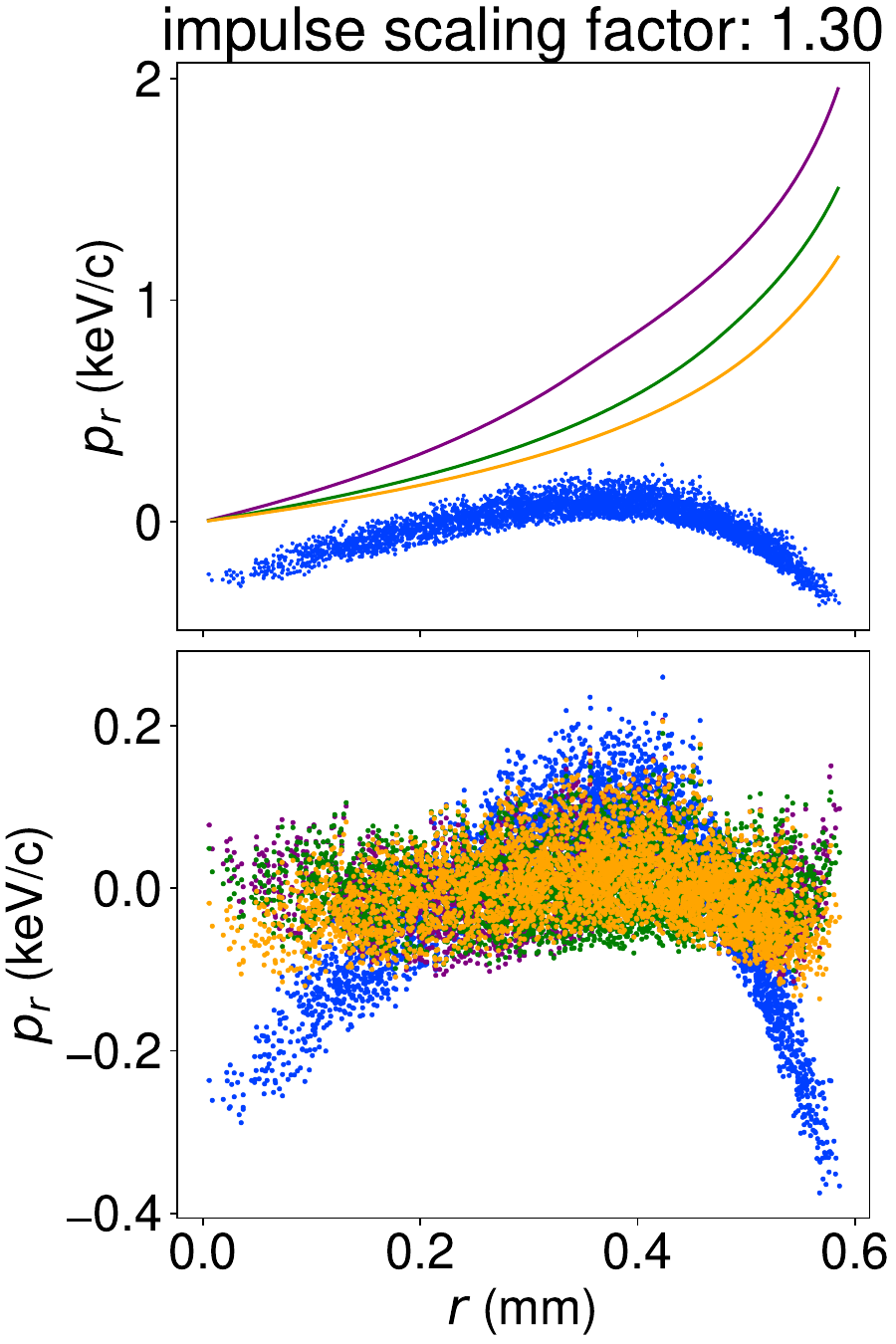}
    \caption{The impulses and phase spaces for all the velocity spreads, using the same scaling factor for all of them. Each color represents the same data as they do in Fig. \ref{fig:analytic_individuals}. The phase spaces for the three cases with different velocity spreads are all roughly straight, despite their optimal scaling factors being different.}
    \label{fig:analytic_combined}
\end{figure}

We can see how much of an effect the scaling factor has by plotting all three of these plots together with the same scaling factor shown in Fig. \ref{fig:analytic_combined}. We see that all of the final phase spaces have, for the most part, been straightened out, despite not being optimal. Indeed the emittances for the low and high velocity spreads have only increased by 2 nm-rad (10 $\%$) from their optimal values.

Based on these results, it is reasonable to assume that the changing velocity spread will not have a large effect on the final shape of the phase space. Changing the velocity spread by a total factor of 4 from the lowest to the highest results in emittances that are within $10\%$ of each other, and the change is caused essentially entirely by a scaling factor on the impulse. 

Therefore, the impulse provided by the collapsing shell is precisely the right shape to undo the curvature caused by space charge in the gun. It provides a larger kick to the radial tails than to the radial center as it collapses, and straightens out the phase space, resulting in an emittance decrease of more than a factor of 2. Recall from the discussion in the introduction that the shell is precisely the wavebroken shock front caused by the strongly nonlinear space charge forces in the gun. Said another way, the phase space nonlinearity in the gun develops such that this final impulse from the collapsing shell compensates it.

\section{The collapsing shell model including space charge}

Now we extend our model to include the slowing of the collapsing shell by the space charge forces of the beam, where we will model the beam as being uniformly distributed. This effect is responsible for the order of magnitude increase in velocity spread during the collapse that was not accounted for in the analytic model, which we rectify in this section.

Let $a(t)$ be the position of a shell particle with initial velocity $v$ as a function of time. The field from the beam at $a$, assuming that $a$ is inside the beam, is given by Gauss's Law:
\begin{equation}
    \vec{E}(a) = \hat{r}\frac{\rho a}{2 \epsilon_0},
\end{equation}
where $\rho$ is the charge density of the beam.
Taking $r_b(t)$ to be the outer radius of the beam as a function of time, which we will determine later, we have
\begin{equation}
    \rho(t) = \frac{\lambda_b}{\pi r_b(t)^2},
\end{equation}
so 
\begin{equation}\label{eq:shell_field}
    \vec{E}(a) = \hat{r} \frac{\lambda_b a}{2\pi \epsilon_0 r_b(t)^2},
\end{equation}
where $\lambda_b$ is the longitudinal charge density of the beam.
Thus, we can write the differential equation
\begin{equation}\label{eq:shell_diff}
    \ddot{a}(t) = \frac{q}{m} \frac{\lambda_b a(t)}{2\pi \epsilon_0 (r_b(t))^2},
\end{equation}
with initial conditions
\begin{equation}
    a(0) = a_0, \, \, \dot{a}(t) = v.
\end{equation}
We now need to determine $r_b$, which is determined by the effects of space charge on the beam itself. We will assume for simplicity that the beam stays as a uniform cylinder as it focuses. We start with the envelope equation in K-V form for an infinitely long, uniform cylinder with radius $r(0) = r_0$:
\begin{equation} \label{eq:envelope}
    \ddot{r}(t) - \frac{\varepsilon^2}{r(t)^3} - \frac{qE(t)}{m} = 0,
\end{equation}
where $\varepsilon$ is the normalized emittance in units of length*momentum.

\begin{figure}
    \centering
    \includegraphics[width=0.6\linewidth]{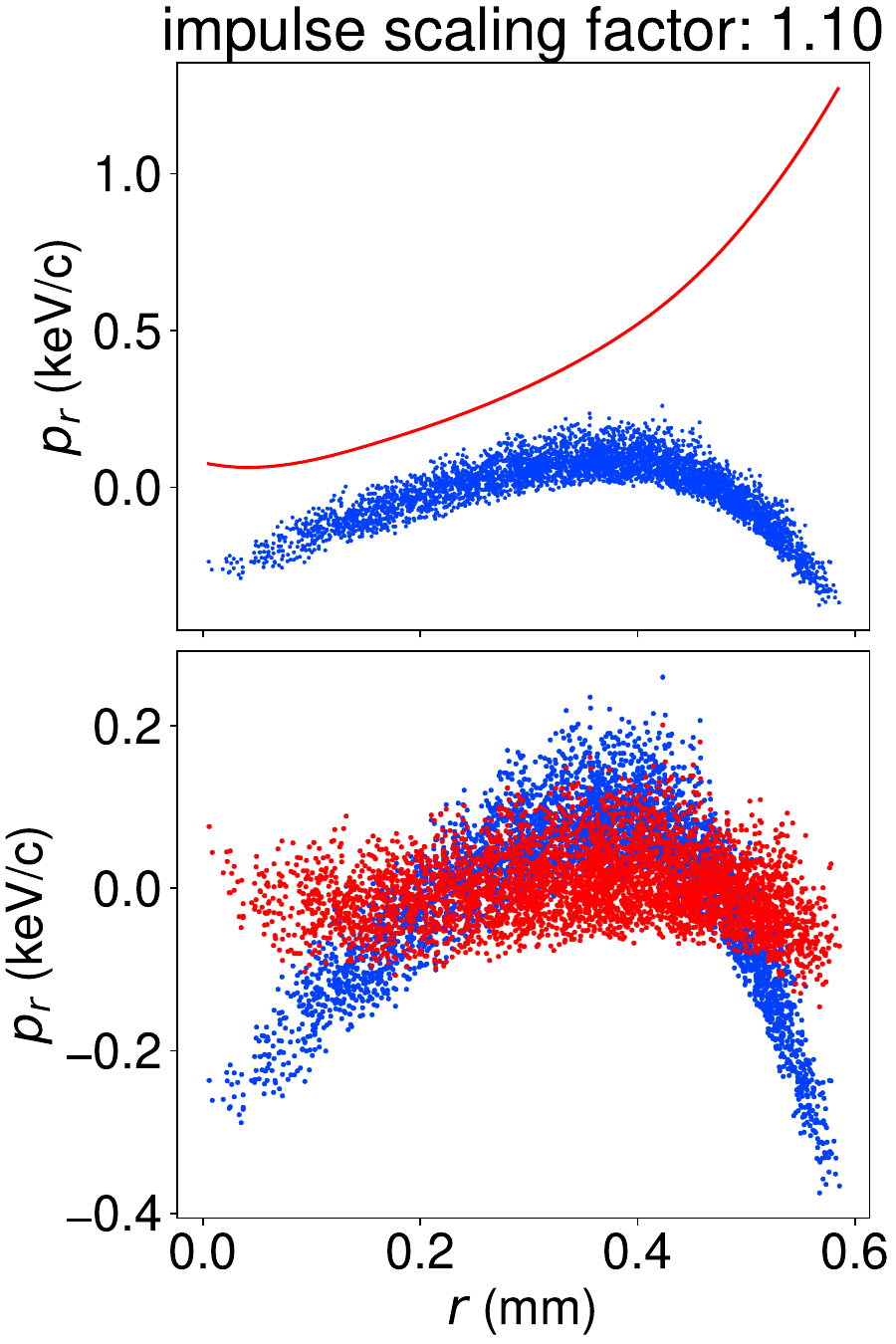}
    \caption{The impulse and final phase space of a collapsing shell with space charge forces taken into account in the dynamics of the shell. We see that, even here, the overall shape is maintained, and appropriate scaling can straighten out the phase space. The final emittance (23.1 nm-rad) is only slightly higher than the best case from before (21.4 nm-rad).}
    \label{fig:space_charge_shell}
\end{figure}
By Gauss's Law, we can compute $E(t)$ as
\begin{equation}
    E(t) = \frac{\rho(t) r(t)}{2 \epsilon_0},
\end{equation}
where $\rho(t)$ is the charge density. We have
\begin{equation} \label{eq:charge_density}
    \rho(t) = \frac{\lambda}{\pi r^2},
\end{equation}
where $\lambda$ is the linear charge density, which remains constant through the process. We can now write the field
\begin{equation}
    E(t) = \frac{\lambda}{2 \pi \epsilon_0 r(t)}.
\end{equation}

Equation \ref{eq:envelope} becomes
\begin{equation}\label{eq:envelope_final}
    \ddot{r}(t) - \frac{\varepsilon^2}{r(t)^3} - \frac{q\lambda}{2 \pi \epsilon_0 m r(t)} = 0.
\end{equation}

Since we are modeling the beam as an infinitely long cylinder, charge that is at a radial position greater than that of the test charge has no effect. Thus, the envelope equation here holds true for all charges inside the cylinder as well, since we can treat all the charge with radial position less than $r$ as an independent beam. Accounting for space charge thus requires replacing the RHS of Eq. \ref{eq:t_p}, the $r_0(1 - Kt)$ in the denominator of Eq. \ref{eq:impulse_int}, and the $r_b(t)$ in Eq. \ref{eq:shell_diff} with the solution of Eq. \ref{eq:envelope_final}.

This becomes intractable analytically, so we solve these equations numerically instead, using the average velocity spread from before. The results are shown in Fig. \ref{fig:space_charge_shell}. Even with the complications introduced from space charge forces affecting the dynamics of the shell collapse, the overall shape of the impulse is maintained. With an appropriate scaling factor, the phase space can be linearized again. With a scale factor of 1.1, the final emittance is only slightly worse than the best case emittance from the previous examples (23.1 nm-rad vs. 21.4 nm-rad). The difference is most likely due to several factors including 3D effects, the initial transverse extent of the shell, and forces applied by the shell after $t_f$.

We can perform a simple back-of-the-envelope calculation to determine the correction from the beam's finite longitudinal extent. The on-axis electric field from a uniform disk of charge with surface charge density $\sigma$ is given by
\begin{equation}
    E_z = \frac{\sigma}{2\epsilon_0}\left(1-\frac{z}{(z^2+a_0^2)^{\frac{1}{2}}}\right).
\end{equation}
We slice the beam longitudinally into disks, each with charge density $\sigma = \frac{\lambda_b d\xi}{\pi a_0^2}$, where $\xi$ is the longitudinal coordinate in the beam's rest frame, with $\xi=0$ defined as the longitudinal center of the beam. The total electric field at a point $\xi=z$ for a uniform cylinder of length $L$ is then given by 
\begin{align}
    E_z(z) = \frac{\lambda_b}{2\pi a_0^2 \epsilon_0}\Bigg[&\int_{-\frac{L}{2}}^z\left(1-\frac{z-\xi}{((z-\xi)^2+a_0^2)^{\frac{1}{2}}}\right)d\xi \\
                                                        - &\int_z^{\frac{L}{2}}\left(1-\frac{\xi-z}{((\xi-z)^2+a_0^2)^{\frac{1}{2}}}\right)d\xi\Bigg]. \nonumber
\end{align}

The contribution to the radial field from the longitudinal field is given by Gauss's Law. Due to superposition, we can calculate the free space effect of the longitudinal field derivative, which can then be added on to the previously calculated radial field as a correction. We have
\begin{align}
\frac{\partial E_z}{\partial z} = \frac{\lambda_b}{2\pi a_0^2 \epsilon_0} \Bigg[2&+\frac{z-\frac{L}{2}}{\left(\left(z-\frac{L}{2}\right)^2 + a_0^2\right)^{\frac{1}{2}}}\\ &- \frac{z+\frac{L}{2}}{\left(\left(z+\frac{L}{2}\right)^2 + a_0^2\right)^{\frac{1}{2}}}\Bigg]\nonumber.
\end{align}
At $z=0$, this becomes
\begin{equation}
    \frac{\partial E_z}{\partial z} = \frac{\lambda_b}{2\pi a_0^2 \epsilon_0}\left(2-\frac{L}{\left(\left(\frac{L}{2}\right)^2 + a_0^2\right)^\frac{1}{2}}\right).
\end{equation}

Since this expression has no $r$ dependence, near the center of the beam, we can approximate the field as being linear in $r$. Comparing this correction coefficient to the expression in Eqn. \ref{eq:shell_field}, we see that the correction is the magnitude of the factor
\begin{equation}
2-\frac{L}{\left(\left(\frac{L}{2}\right)^2 + a_0^2\right)^\frac{1}{2}} .
\end{equation}
With our beam parameters, this represents a percent-level correction to the space charge field.

Therefore, we see that a simple model of an infinitely long cylindrical shell collapsing into an infinitely long cylindrical beam produces the correct shape to fix the phase space curvature, even after accounting for complications such as velocity spread and space charge dynamics.

\section{Summary}

In summary, we have demonstrated in simulation a novel approach to nonlinear slice emittance compensation, utilizing the wavebreaking characteristic of  nonlinear space charge forces in dense beams to undo the emittance degradation caused by those very same nonlinear forces on the core of the beam. The wavebroken shock front at the edge of the beam is nonlaminarly focused, crashing through the core of the beam and linearizing it in the process. We have found that in simulated photoinjectors utilizing cutting-edge low MTE photocathodes, the employment of this wave-crashing linearization can reduce the emittance by more than a factor of 2.

Furthermore, we have constructed a simple analytic model of the wave-crashing process which agrees well with simulation, showing that the curvature of the linearizing force arises from the large velocity spread inherent to the wavebroken shock front. The extreme nonlinearity at the edge of the beam, instead of being an undesired artifact, becomes critical to the linearization of the core.

The charge forming this nonlinearity can subsequently be discarded by an aperture. While the presence of a physical aperture can exhibit wakefield effects which are beyond the scope of this paper and are the subject of future study, we note that in some applications a physical aperture may not be required. In certain applications of high brightness beams, such as FELs or inverse Compton scattering, the sacrifical charge may not contribute meaningfully to the radiation production and might be ignored without the necessity of physically discarding it. 

As higher brightnesses become necessary to push the performance of the next generation of FELs and colliders, the commensurate charge densities at the cathode demand novel techniques of mitigating space charge-induced emittance degradation. We have demonstrated one such technique, providing a potential path to taking full advantage of the next generation of rf guns and photocathode.

\begin{acknowledgements}
This work was supported by the U.S Department of
Energy, Grant No. DE-SC0020144, and U.S. National
Science Foundation Grant No. PHY-1549132, the Center
for Bright Beams. 
\end{acknowledgements}

\bibliography{main.bib}

\begin{thebibliography}{26}%
\makeatletter
\providecommand \@ifxundefined [1]{%
 \@ifx{#1\undefined}
}%
\providecommand \@ifnum [1]{%
 \ifnum #1\expandafter \@firstoftwo
 \else \expandafter \@secondoftwo
 \fi
}%
\providecommand \@ifx [1]{%
 \ifx #1\expandafter \@firstoftwo
 \else \expandafter \@secondoftwo
 \fi
}%
\providecommand \natexlab [1]{#1}%
\providecommand \enquote  [1]{``#1''}%
\providecommand \bibnamefont  [1]{#1}%
\providecommand \bibfnamefont [1]{#1}%
\providecommand \citenamefont [1]{#1}%
\providecommand \href@noop [0]{\@secondoftwo}%
\providecommand \href [0]{\begingroup \@sanitize@url \@href}%
\providecommand \@href[1]{\@@startlink{#1}\@@href}%
\providecommand \@@href[1]{\endgroup#1\@@endlink}%
\providecommand \@sanitize@url [0]{\catcode `\\12\catcode `\$12\catcode
  `\&12\catcode `\#12\catcode `\^12\catcode `\_12\catcode `\%12\relax}%
\providecommand \@@startlink[1]{}%
\providecommand \@@endlink[0]{}%
\providecommand \url  [0]{\begingroup\@sanitize@url \@url }%
\providecommand \@url [1]{\endgroup\@href {#1}{\urlprefix }}%
\providecommand \urlprefix  [0]{URL }%
\providecommand \Eprint [0]{\href }%
\providecommand \doibase [0]{https://doi.org/}%
\providecommand \selectlanguage [0]{\@gobble}%
\providecommand \bibinfo  [0]{\@secondoftwo}%
\providecommand \bibfield  [0]{\@secondoftwo}%
\providecommand \translation [1]{[#1]}%
\providecommand \BibitemOpen [0]{}%
\providecommand \bibitemStop [0]{}%
\providecommand \bibitemNoStop [0]{.\EOS\space}%
\providecommand \EOS [0]{\spacefactor3000\relax}%
\providecommand \BibitemShut  [1]{\csname bibitem#1\endcsname}%
\let\auto@bib@innerbib\@empty
\bibitem [{\citenamefont {Serafini}\ and\ \citenamefont
  {Rosenzweig}(1997)}]{serafiniEnvelopeAnalysisIntense1997}%
  \BibitemOpen
  \bibfield  {author} {\bibinfo {author} {\bibfnamefont {L.}~\bibnamefont
  {Serafini}}\ and\ \bibinfo {author} {\bibfnamefont {J.~B.}\ \bibnamefont
  {Rosenzweig}},\ }\bibfield  {title} {\bibinfo {title} {Envelope analysis of
  intense relativistic quasilaminar beams in rf photoinjectors:{{mA}} theory of
  emittance compensation},\ }\href {https://doi.org/10.1103/PhysRevE.55.7565}
  {\bibfield  {journal} {\bibinfo  {journal} {Phys. Rev. E}\ }\textbf {\bibinfo
  {volume} {55}},\ \bibinfo {pages} {7565} (\bibinfo {year}
  {1997})}\BibitemShut {NoStop}%
\bibitem [{\citenamefont
  {Carlsten}(1989)}]{carlstenNewPhotoelectricInjector1989}%
  \BibitemOpen
  \bibfield  {author} {\bibinfo {author} {\bibfnamefont {B.~E.}\ \bibnamefont
  {Carlsten}},\ }\bibfield  {title} {\bibinfo {title} {New photoelectric
  injector design for the los alamos national laboratory {{XUV FEL}}
  accelerator},\ }\href {https://doi.org/10.1016/0168-9002(89)90472-5}
  {\bibfield  {journal} {\bibinfo  {journal} {Nucl. Instrum. Meth. A}\ }\textbf
  {\bibinfo {volume} {285}},\ \bibinfo {pages} {313} (\bibinfo {year}
  {1989})}\BibitemShut {NoStop}%
\bibitem [{\citenamefont {Chen}\ \emph {et~al.}(2021)\citenamefont {Chen},
  \citenamefont {Zheng}, \citenamefont {Huang}, \citenamefont {Song},
  \citenamefont {Du}, \citenamefont {Li}, \citenamefont {Huang},\ and\
  \citenamefont {Tang}}]{chenAnalysisSliceTransverse2021}%
  \BibitemOpen
  \bibfield  {author} {\bibinfo {author} {\bibfnamefont {H.}~\bibnamefont
  {Chen}}, \bibinfo {author} {\bibfnamefont {L.}~\bibnamefont {Zheng}},
  \bibinfo {author} {\bibfnamefont {P.}~\bibnamefont {Huang}}, \bibinfo
  {author} {\bibfnamefont {C.}~\bibnamefont {Song}}, \bibinfo {author}
  {\bibfnamefont {Y.}~\bibnamefont {Du}}, \bibinfo {author} {\bibfnamefont
  {R.}~\bibnamefont {Li}}, \bibinfo {author} {\bibfnamefont {W.}~\bibnamefont
  {Huang}},\ and\ \bibinfo {author} {\bibfnamefont {C.}~\bibnamefont {Tang}},\
  }\bibfield  {title} {\bibinfo {title} {Analysis of slice transverse emittance
  evolution in a very-high-frequency gun photoinjector},\ }\href
  {https://doi.org/10.1103/PhysRevAccelBeams.24.124402} {\bibfield  {journal}
  {\bibinfo  {journal} {Phys. Rev. Accel. Beams}\ }\textbf {\bibinfo {volume}
  {24}},\ \bibinfo {pages} {124402} (\bibinfo {year} {2021})}\BibitemShut
  {NoStop}%
\bibitem [{\citenamefont {Vecchione}\ \emph {et~al.}(2011)\citenamefont
  {Vecchione}, \citenamefont {{Ben-Zvi}}, \citenamefont {Dowell}, \citenamefont
  {Feng}, \citenamefont {Rao}, \citenamefont {Smedley}, \citenamefont {Wan},\
  and\ \citenamefont {Padmore}}]{vecchioneLowEmittanceHigh2011}%
  \BibitemOpen
  \bibfield  {author} {\bibinfo {author} {\bibfnamefont {T.}~\bibnamefont
  {Vecchione}}, \bibinfo {author} {\bibfnamefont {I.}~\bibnamefont
  {{Ben-Zvi}}}, \bibinfo {author} {\bibfnamefont {D.~H.}\ \bibnamefont
  {Dowell}}, \bibinfo {author} {\bibfnamefont {J.}~\bibnamefont {Feng}},
  \bibinfo {author} {\bibfnamefont {T.}~\bibnamefont {Rao}}, \bibinfo {author}
  {\bibfnamefont {J.}~\bibnamefont {Smedley}}, \bibinfo {author} {\bibfnamefont
  {W.}~\bibnamefont {Wan}},\ and\ \bibinfo {author} {\bibfnamefont {H.~A.}\
  \bibnamefont {Padmore}},\ }\bibfield  {title} {\bibinfo {title} {A low
  emittance and high efficiency visible light photocathode for high brightness
  accelerator-based {{X-ray}} light sources},\ }\href
  {https://doi.org/10.1063/1.3612916} {\bibfield  {journal} {\bibinfo
  {journal} {Applied Physics Letters}\ }\textbf {\bibinfo {volume} {99}},\
  \bibinfo {pages} {034103} (\bibinfo {year} {2011})}\BibitemShut {NoStop}%
\bibitem [{\citenamefont {Feng}\ \emph {et~al.}(2015)\citenamefont {Feng},
  \citenamefont {Nasiatka}, \citenamefont {Wan}, \citenamefont {Karkare},
  \citenamefont {Smedley},\ and\ \citenamefont
  {Padmore}}]{fengThermalLimitIntrinsic2015}%
  \BibitemOpen
  \bibfield  {author} {\bibinfo {author} {\bibfnamefont {J.}~\bibnamefont
  {Feng}}, \bibinfo {author} {\bibfnamefont {J.}~\bibnamefont {Nasiatka}},
  \bibinfo {author} {\bibfnamefont {W.}~\bibnamefont {Wan}}, \bibinfo {author}
  {\bibfnamefont {S.}~\bibnamefont {Karkare}}, \bibinfo {author} {\bibfnamefont
  {J.}~\bibnamefont {Smedley}},\ and\ \bibinfo {author} {\bibfnamefont {H.~A.}\
  \bibnamefont {Padmore}},\ }\bibfield  {title} {\bibinfo {title} {Thermal
  limit to the intrinsic emittance from metal photocathodes},\ }\href
  {https://doi.org/10.1063/1.4931976} {\bibfield  {journal} {\bibinfo
  {journal} {Applied Physics Letters}\ }\textbf {\bibinfo {volume} {107}},\
  \bibinfo {pages} {134101} (\bibinfo {year} {2015})}\BibitemShut {NoStop}%
\bibitem [{\citenamefont {Anderson}\ and\ \citenamefont
  {Rosenzweig}(2000)}]{andersonNonequilibriumTransverseMotion2000}%
  \BibitemOpen
  \bibfield  {author} {\bibinfo {author} {\bibfnamefont {S.~G.}\ \bibnamefont
  {Anderson}}\ and\ \bibinfo {author} {\bibfnamefont {J.~B.}\ \bibnamefont
  {Rosenzweig}},\ }\bibfield  {title} {\bibinfo {title} {Nonequilibrium
  transverse motion and emittance growth in ultrarelativistic space-charge
  dominated beams},\ }\href {https://doi.org/10.1103/PhysRevSTAB.3.094201}
  {\bibfield  {journal} {\bibinfo  {journal} {Phys. Rev. ST Accel. Beams}\
  }\textbf {\bibinfo {volume} {3}},\ \bibinfo {pages} {094201} (\bibinfo {year}
  {2000})}\BibitemShut {NoStop}%
\bibitem [{\citenamefont
  {Floettmann}(2017)}]{floettmannEmittanceCompensationSplit2017}%
  \BibitemOpen
  \bibfield  {author} {\bibinfo {author} {\bibfnamefont {K.}~\bibnamefont
  {Floettmann}},\ }\bibfield  {title} {\bibinfo {title} {Emittance compensation
  in split photoinjectors},\ }\href
  {https://doi.org/10.1103/PhysRevAccelBeams.20.013401} {\bibfield  {journal}
  {\bibinfo  {journal} {Phys. Rev. Accel. Beams}\ }\textbf {\bibinfo {volume}
  {20}},\ \bibinfo {pages} {013401} (\bibinfo {year} {2017})}\BibitemShut
  {NoStop}%
\bibitem [{\citenamefont {Li}\ \emph {et~al.}(2012)\citenamefont {Li},
  \citenamefont {Roberts}, \citenamefont {Scoby}, \citenamefont {To},\ and\
  \citenamefont {Musumeci}}]{liNanometerEmittanceUltralow2012}%
  \BibitemOpen
  \bibfield  {author} {\bibinfo {author} {\bibfnamefont {R.~K.}\ \bibnamefont
  {Li}}, \bibinfo {author} {\bibfnamefont {K.~G.}\ \bibnamefont {Roberts}},
  \bibinfo {author} {\bibfnamefont {C.~M.}\ \bibnamefont {Scoby}}, \bibinfo
  {author} {\bibfnamefont {H.}~\bibnamefont {To}},\ and\ \bibinfo {author}
  {\bibfnamefont {P.}~\bibnamefont {Musumeci}},\ }\bibfield  {title} {\bibinfo
  {title} {Nanometer emittance ultralow charge beams from rf photoinjectors},\
  }\href {https://doi.org/10.1103/PhysRevSTAB.15.090702} {\bibfield  {journal}
  {\bibinfo  {journal} {Phys. Rev. ST Accel. Beams}\ }\textbf {\bibinfo
  {volume} {15}},\ \bibinfo {pages} {090702} (\bibinfo {year}
  {2012})}\BibitemShut {NoStop}%
\bibitem [{\citenamefont {Zerbe}\ \emph {et~al.}(2018)\citenamefont {Zerbe},
  \citenamefont {Xiang}, \citenamefont {Ruan}, \citenamefont {Lund},\ and\
  \citenamefont {Duxbury}}]{zerbeDynamicalBunchingDensity2018}%
  \BibitemOpen
  \bibfield  {author} {\bibinfo {author} {\bibfnamefont {B.~S.}\ \bibnamefont
  {Zerbe}}, \bibinfo {author} {\bibfnamefont {X.}~\bibnamefont {Xiang}},
  \bibinfo {author} {\bibfnamefont {C.-Y.}\ \bibnamefont {Ruan}}, \bibinfo
  {author} {\bibfnamefont {S.}~\bibnamefont {Lund}},\ and\ \bibinfo {author}
  {\bibfnamefont {P.}~\bibnamefont {Duxbury}},\ }\bibfield  {title} {\bibinfo
  {title} {Dynamical bunching and density peaks in expanding {{Coulomb}}
  clouds},\ }\bibfield  {journal} {\bibinfo  {journal} {Phys. Rev. Accel.
  Beams}\ }\textbf {\bibinfo {volume} {21}},\ \href
  {https://doi.org/10.1103/physrevaccelbeams.21.064201}
  {10.1103/physrevaccelbeams.21.064201} (\bibinfo {year} {2018})\BibitemShut
  {NoStop}%
\bibitem [{\citenamefont {Bazarov}\ \emph {et~al.}(2009)\citenamefont
  {Bazarov}, \citenamefont {Dunham},\ and\ \citenamefont
  {Sinclair}}]{bazarovMaximumAchievableBeam2009}%
  \BibitemOpen
  \bibfield  {author} {\bibinfo {author} {\bibfnamefont {I.~V.}\ \bibnamefont
  {Bazarov}}, \bibinfo {author} {\bibfnamefont {B.~M.}\ \bibnamefont
  {Dunham}},\ and\ \bibinfo {author} {\bibfnamefont {C.~K.}\ \bibnamefont
  {Sinclair}},\ }\bibfield  {title} {\bibinfo {title} {Maximum {{Achievable
  Beam Brightness}} from {{Photoinjectors}}},\ }\href
  {https://doi.org/10.1103/PhysRevLett.102.104801} {\bibfield  {journal}
  {\bibinfo  {journal} {Phys. Rev. Lett.}\ }\textbf {\bibinfo {volume} {102}},\
  \bibinfo {pages} {104801} (\bibinfo {year} {2009})}\BibitemShut {NoStop}%
\bibitem [{\citenamefont {Rosenzweig}\ \emph {et~al.}(2020)\citenamefont
  {Rosenzweig}, \citenamefont {Majernik}, \citenamefont {Robles}, \citenamefont
  {Andonian}, \citenamefont {Camacho}, \citenamefont {Fukasawa}, \citenamefont
  {Kogar}, \citenamefont {Lawler}, \citenamefont {Miao}, \citenamefont
  {Musumeci}, \citenamefont {Naranjo}, \citenamefont {Sakai}, \citenamefont
  {Candler}, \citenamefont {Pound}, \citenamefont {Pellegrini}, \citenamefont
  {Emma}, \citenamefont {Halavanau}, \citenamefont {Hastings}, \citenamefont
  {Li}, \citenamefont {Nasr}, \citenamefont {Tantawi}, \citenamefont
  {Anisimov}, \citenamefont {Carlsten}, \citenamefont {Krawczyk}, \citenamefont
  {Simakov}, \citenamefont {Faillace}, \citenamefont {Ferrario}, \citenamefont
  {Spataro}, \citenamefont {Karkare}, \citenamefont {Maxson}, \citenamefont
  {Ma}, \citenamefont {Wurtele}, \citenamefont {Murokh}, \citenamefont
  {Zholents}, \citenamefont {Cianchi}, \citenamefont {Cocco},\ and\
  \citenamefont {van~der Geer}}]{rosenzweigUltracompactXrayFreeelectron2020}%
  \BibitemOpen
  \bibfield  {author} {\bibinfo {author} {\bibfnamefont {J.~B.}\ \bibnamefont
  {Rosenzweig}}, \bibinfo {author} {\bibfnamefont {N.}~\bibnamefont
  {Majernik}}, \bibinfo {author} {\bibfnamefont {R.~R.}\ \bibnamefont
  {Robles}}, \bibinfo {author} {\bibfnamefont {G.}~\bibnamefont {Andonian}},
  \bibinfo {author} {\bibfnamefont {O.}~\bibnamefont {Camacho}}, \bibinfo
  {author} {\bibfnamefont {A.}~\bibnamefont {Fukasawa}}, \bibinfo {author}
  {\bibfnamefont {A.}~\bibnamefont {Kogar}}, \bibinfo {author} {\bibfnamefont
  {G.}~\bibnamefont {Lawler}}, \bibinfo {author} {\bibfnamefont
  {J.}~\bibnamefont {Miao}}, \bibinfo {author} {\bibfnamefont {P.}~\bibnamefont
  {Musumeci}}, \bibinfo {author} {\bibfnamefont {B.}~\bibnamefont {Naranjo}},
  \bibinfo {author} {\bibfnamefont {Y.}~\bibnamefont {Sakai}}, \bibinfo
  {author} {\bibfnamefont {R.}~\bibnamefont {Candler}}, \bibinfo {author}
  {\bibfnamefont {B.}~\bibnamefont {Pound}}, \bibinfo {author} {\bibfnamefont
  {C.}~\bibnamefont {Pellegrini}}, \bibinfo {author} {\bibfnamefont
  {C.}~\bibnamefont {Emma}}, \bibinfo {author} {\bibfnamefont {A.}~\bibnamefont
  {Halavanau}}, \bibinfo {author} {\bibfnamefont {J.}~\bibnamefont {Hastings}},
  \bibinfo {author} {\bibfnamefont {Z.}~\bibnamefont {Li}}, \bibinfo {author}
  {\bibfnamefont {M.}~\bibnamefont {Nasr}}, \bibinfo {author} {\bibfnamefont
  {S.}~\bibnamefont {Tantawi}}, \bibinfo {author} {\bibfnamefont
  {P.}~\bibnamefont {Anisimov}}, \bibinfo {author} {\bibfnamefont
  {B.}~\bibnamefont {Carlsten}}, \bibinfo {author} {\bibfnamefont
  {F.}~\bibnamefont {Krawczyk}}, \bibinfo {author} {\bibfnamefont
  {E.}~\bibnamefont {Simakov}}, \bibinfo {author} {\bibfnamefont
  {L.}~\bibnamefont {Faillace}}, \bibinfo {author} {\bibfnamefont
  {M.}~\bibnamefont {Ferrario}}, \bibinfo {author} {\bibfnamefont
  {B.}~\bibnamefont {Spataro}}, \bibinfo {author} {\bibfnamefont
  {S.}~\bibnamefont {Karkare}}, \bibinfo {author} {\bibfnamefont
  {J.}~\bibnamefont {Maxson}}, \bibinfo {author} {\bibfnamefont
  {Y.}~\bibnamefont {Ma}}, \bibinfo {author} {\bibfnamefont {J.}~\bibnamefont
  {Wurtele}}, \bibinfo {author} {\bibfnamefont {A.}~\bibnamefont {Murokh}},
  \bibinfo {author} {\bibfnamefont {A.}~\bibnamefont {Zholents}}, \bibinfo
  {author} {\bibfnamefont {A.}~\bibnamefont {Cianchi}}, \bibinfo {author}
  {\bibfnamefont {D.}~\bibnamefont {Cocco}},\ and\ \bibinfo {author}
  {\bibfnamefont {S.~B.}\ \bibnamefont {van~der Geer}},\ }\bibfield  {title}
  {\bibinfo {title} {An ultra-compact x-ray free-electron laser},\ }\href
  {https://doi.org/10.1088/1367-2630/abb16c} {\bibfield  {journal} {\bibinfo
  {journal} {New J. Phys.}\ }\textbf {\bibinfo {volume} {22}},\ \bibinfo
  {pages} {093067} (\bibinfo {year} {2020})}\BibitemShut {NoStop}%
\bibitem [{\citenamefont {Tantawi}\ \emph {et~al.}(2020)\citenamefont
  {Tantawi}, \citenamefont {Nasr}, \citenamefont {Li}, \citenamefont
  {Limborg},\ and\ \citenamefont
  {Borchard}}]{tantawiDesignDemonstrationDistributedcoupling2020}%
  \BibitemOpen
  \bibfield  {author} {\bibinfo {author} {\bibfnamefont {S.}~\bibnamefont
  {Tantawi}}, \bibinfo {author} {\bibfnamefont {M.}~\bibnamefont {Nasr}},
  \bibinfo {author} {\bibfnamefont {Z.}~\bibnamefont {Li}}, \bibinfo {author}
  {\bibfnamefont {C.}~\bibnamefont {Limborg}},\ and\ \bibinfo {author}
  {\bibfnamefont {P.}~\bibnamefont {Borchard}},\ }\bibfield  {title} {\bibinfo
  {title} {Design and demonstration of a distributed-coupling linear
  accelerator structure},\ }\href
  {https://doi.org/10.1103/PhysRevAccelBeams.23.092001} {\bibfield  {journal}
  {\bibinfo  {journal} {Phys. Rev. Accel. Beams}\ }\textbf {\bibinfo {volume}
  {23}},\ \bibinfo {pages} {092001} (\bibinfo {year} {2020})}\BibitemShut
  {NoStop}%
\bibitem [{\citenamefont {Rosenzweig}\ \emph {et~al.}(2018)\citenamefont
  {Rosenzweig}, \citenamefont {Cahill}, \citenamefont {Carlsten}, \citenamefont
  {Castorina}, \citenamefont {Croia}, \citenamefont {Emma}, \citenamefont
  {Fukusawa}, \citenamefont {Spataro}, \citenamefont {Alesini}, \citenamefont
  {Dolgashev}, \citenamefont {Ferrario}, \citenamefont {Lawler}, \citenamefont
  {Li}, \citenamefont {Limborg}, \citenamefont {Maxson}, \citenamefont
  {Musumeci}, \citenamefont {Pompili}, \citenamefont {Tantawi},\ and\
  \citenamefont {Williams}}]{rosenzweigUltrahighBrightnessElectron2018}%
  \BibitemOpen
  \bibfield  {author} {\bibinfo {author} {\bibfnamefont {J.~B.}\ \bibnamefont
  {Rosenzweig}}, \bibinfo {author} {\bibfnamefont {A.}~\bibnamefont {Cahill}},
  \bibinfo {author} {\bibfnamefont {B.}~\bibnamefont {Carlsten}}, \bibinfo
  {author} {\bibfnamefont {G.}~\bibnamefont {Castorina}}, \bibinfo {author}
  {\bibfnamefont {M.}~\bibnamefont {Croia}}, \bibinfo {author} {\bibfnamefont
  {C.}~\bibnamefont {Emma}}, \bibinfo {author} {\bibfnamefont {A.}~\bibnamefont
  {Fukusawa}}, \bibinfo {author} {\bibfnamefont {B.}~\bibnamefont {Spataro}},
  \bibinfo {author} {\bibfnamefont {D.}~\bibnamefont {Alesini}}, \bibinfo
  {author} {\bibfnamefont {V.}~\bibnamefont {Dolgashev}}, \bibinfo {author}
  {\bibfnamefont {M.}~\bibnamefont {Ferrario}}, \bibinfo {author}
  {\bibfnamefont {G.}~\bibnamefont {Lawler}}, \bibinfo {author} {\bibfnamefont
  {R.}~\bibnamefont {Li}}, \bibinfo {author} {\bibfnamefont {C.}~\bibnamefont
  {Limborg}}, \bibinfo {author} {\bibfnamefont {J.}~\bibnamefont {Maxson}},
  \bibinfo {author} {\bibfnamefont {P.}~\bibnamefont {Musumeci}}, \bibinfo
  {author} {\bibfnamefont {R.}~\bibnamefont {Pompili}}, \bibinfo {author}
  {\bibfnamefont {S.}~\bibnamefont {Tantawi}},\ and\ \bibinfo {author}
  {\bibfnamefont {O.}~\bibnamefont {Williams}},\ }\bibfield  {title} {\bibinfo
  {title} {Ultra-high brightness electron beams from very-high field cryogenic
  radiofrequency photocathode sources},\ }\href
  {https://doi.org/10.1016/j.nima.2018.01.061} {\bibfield  {journal} {\bibinfo
  {journal} {Nuclear Instruments and Methods in Physics Research Section A:
  Accelerators, Spectrometers, Detectors and Associated Equipment}\ }\bibinfo
  {series} {3rd {{European Advanced Accelerator Concepts}} Workshop
  ({{EAAC2017}})},\ \textbf {\bibinfo {volume} {909}},\ \bibinfo {pages} {224}
  (\bibinfo {year} {2018})}\BibitemShut {NoStop}%
\bibitem [{\citenamefont {Cahill}\ \emph {et~al.}(2018)\citenamefont {Cahill},
  \citenamefont {Rosenzweig}, \citenamefont {Dolgashev}, \citenamefont
  {Tantawi},\ and\ \citenamefont
  {Weathersby}}]{cahillHighGradientExperiments2018}%
  \BibitemOpen
  \bibfield  {author} {\bibinfo {author} {\bibfnamefont {A.~D.}\ \bibnamefont
  {Cahill}}, \bibinfo {author} {\bibfnamefont {J.~B.}\ \bibnamefont
  {Rosenzweig}}, \bibinfo {author} {\bibfnamefont {V.~A.}\ \bibnamefont
  {Dolgashev}}, \bibinfo {author} {\bibfnamefont {S.~G.}\ \bibnamefont
  {Tantawi}},\ and\ \bibinfo {author} {\bibfnamefont {S.}~\bibnamefont
  {Weathersby}},\ }\bibfield  {title} {\bibinfo {title} {High gradient
  experiments with \${{X}}\$-band cryogenic copper accelerating cavities},\
  }\href {https://doi.org/10.1103/PhysRevAccelBeams.21.102002} {\bibfield
  {journal} {\bibinfo  {journal} {Phys. Rev. Accel. Beams}\ }\textbf {\bibinfo
  {volume} {21}},\ \bibinfo {pages} {102002} (\bibinfo {year}
  {2018})}\BibitemShut {NoStop}%
\bibitem [{\citenamefont {Filippetto}\ \emph {et~al.}(2014)\citenamefont
  {Filippetto}, \citenamefont {Musumeci}, \citenamefont {Zolotorev},\ and\
  \citenamefont {Stupakov}}]{filippettoMaximumCurrentDensity2014}%
  \BibitemOpen
  \bibfield  {author} {\bibinfo {author} {\bibfnamefont {D.}~\bibnamefont
  {Filippetto}}, \bibinfo {author} {\bibfnamefont {P.}~\bibnamefont
  {Musumeci}}, \bibinfo {author} {\bibfnamefont {M.}~\bibnamefont
  {Zolotorev}},\ and\ \bibinfo {author} {\bibfnamefont {G.}~\bibnamefont
  {Stupakov}},\ }\bibfield  {title} {\bibinfo {title} {Maximum current density
  and beam brightness achievable by laser-driven electron sources},\ }\bibfield
   {journal} {\bibinfo  {journal} {Phys. Rev. ST Accel. Beams}\ }\textbf
  {\bibinfo {volume} {17}},\ \href
  {https://doi.org/10.1103/physrevstab.17.024201}
  {10.1103/physrevstab.17.024201} (\bibinfo {year} {2014})\BibitemShut
  {NoStop}%
\bibitem [{\citenamefont {Akre}\ \emph {et~al.}(2008)\citenamefont {Akre},
  \citenamefont {Dowell}, \citenamefont {Emma}, \citenamefont {Frisch},
  \citenamefont {Gilevich}, \citenamefont {Hays}, \citenamefont {Hering},
  \citenamefont {Iverson}, \citenamefont {{Limborg-Deprey}}, \citenamefont
  {Loos}, \citenamefont {Miahnahri}, \citenamefont {Schmerge}, \citenamefont
  {Turner}, \citenamefont {Welch}, \citenamefont {White},\ and\ \citenamefont
  {Wu}}]{akreCommissioningLinacCoherent2008}%
  \BibitemOpen
  \bibfield  {author} {\bibinfo {author} {\bibfnamefont {R.}~\bibnamefont
  {Akre}}, \bibinfo {author} {\bibfnamefont {D.}~\bibnamefont {Dowell}},
  \bibinfo {author} {\bibfnamefont {P.}~\bibnamefont {Emma}}, \bibinfo {author}
  {\bibfnamefont {J.}~\bibnamefont {Frisch}}, \bibinfo {author} {\bibfnamefont
  {S.}~\bibnamefont {Gilevich}}, \bibinfo {author} {\bibfnamefont
  {G.}~\bibnamefont {Hays}}, \bibinfo {author} {\bibfnamefont {{\relax
  Ph}.}~\bibnamefont {Hering}}, \bibinfo {author} {\bibfnamefont
  {R.}~\bibnamefont {Iverson}}, \bibinfo {author} {\bibfnamefont
  {C.}~\bibnamefont {{Limborg-Deprey}}}, \bibinfo {author} {\bibfnamefont
  {H.}~\bibnamefont {Loos}}, \bibinfo {author} {\bibfnamefont {A.}~\bibnamefont
  {Miahnahri}}, \bibinfo {author} {\bibfnamefont {J.}~\bibnamefont {Schmerge}},
  \bibinfo {author} {\bibfnamefont {J.}~\bibnamefont {Turner}}, \bibinfo
  {author} {\bibfnamefont {J.}~\bibnamefont {Welch}}, \bibinfo {author}
  {\bibfnamefont {W.}~\bibnamefont {White}},\ and\ \bibinfo {author}
  {\bibfnamefont {J.}~\bibnamefont {Wu}},\ }\bibfield  {title} {\bibinfo
  {title} {Commissioning the {{Linac Coherent Light Source}} injector},\ }\href
  {https://doi.org/10.1103/PhysRevSTAB.11.030703} {\bibfield  {journal}
  {\bibinfo  {journal} {Phys. Rev. ST Accel. Beams}\ }\textbf {\bibinfo
  {volume} {11}},\ \bibinfo {pages} {030703} (\bibinfo {year}
  {2008})}\BibitemShut {NoStop}%
\bibitem [{\citenamefont {Shu}\ \emph {et~al.}(2019)\citenamefont {Shu},
  \citenamefont {Chen}, \citenamefont {Lal}, \citenamefont {Qian},
  \citenamefont {Shaker},\ and\ \citenamefont
  {Stephan}}]{shuFIRSTDESIGNSTUDIES2019}%
  \BibitemOpen
  \bibfield  {author} {\bibinfo {author} {\bibfnamefont {G.}~\bibnamefont
  {Shu}}, \bibinfo {author} {\bibfnamefont {Y.}~\bibnamefont {Chen}}, \bibinfo
  {author} {\bibfnamefont {S.}~\bibnamefont {Lal}}, \bibinfo {author}
  {\bibfnamefont {H.}~\bibnamefont {Qian}}, \bibinfo {author} {\bibfnamefont
  {S.~H.}\ \bibnamefont {Shaker}},\ and\ \bibinfo {author} {\bibfnamefont
  {F.}~\bibnamefont {Stephan}},\ }\bibfield  {title} {\bibinfo {title} {{{FIRST
  DESIGN STUDIES OF A NC CW RF GUN FOR EUROPEAN XFEL}}},\ }\bibfield  {journal}
  {\bibinfo  {journal} {Proceedings of the 10th Int. Particle Accelerator
  Conf.}\ }\textbf {\bibinfo {volume} {IPAC2019}},\ \href
  {https://doi.org/10.18429/JACOW-IPAC2019-TUPRB010}
  {10.18429/JACOW-IPAC2019-TUPRB010} (\bibinfo {year} {2019})\BibitemShut
  {NoStop}%
\bibitem [{\citenamefont {Karkare}\ \emph {et~al.}(2020)\citenamefont
  {Karkare}, \citenamefont {Adhikari}, \citenamefont {Schroeder}, \citenamefont
  {Nangoi}, \citenamefont {Arias}, \citenamefont {Maxson},\ and\ \citenamefont
  {Padmore}}]{karkareUltracoldElectronsNearthreshold2020}%
  \BibitemOpen
  \bibfield  {author} {\bibinfo {author} {\bibfnamefont {S.}~\bibnamefont
  {Karkare}}, \bibinfo {author} {\bibfnamefont {G.}~\bibnamefont {Adhikari}},
  \bibinfo {author} {\bibfnamefont {W.~A.}\ \bibnamefont {Schroeder}}, \bibinfo
  {author} {\bibfnamefont {J.~K.}\ \bibnamefont {Nangoi}}, \bibinfo {author}
  {\bibfnamefont {T.}~\bibnamefont {Arias}}, \bibinfo {author} {\bibfnamefont
  {J.}~\bibnamefont {Maxson}},\ and\ \bibinfo {author} {\bibfnamefont
  {H.}~\bibnamefont {Padmore}},\ }\bibfield  {title} {\bibinfo {title}
  {Ultracold electrons via near-threshold photoemission from single-crystal
  {{Cu}} (100)},\ }\href@noop {} {\bibfield  {journal} {\bibinfo  {journal}
  {Physical review letters}\ }\textbf {\bibinfo {volume} {125}},\ \bibinfo
  {pages} {054801} (\bibinfo {year} {2020})}\BibitemShut {NoStop}%
\bibitem [{\citenamefont {Bartnik}\ \emph {et~al.}(2015)\citenamefont
  {Bartnik}, \citenamefont {Gulliford}, \citenamefont {Bazarov}, \citenamefont
  {Cultera},\ and\ \citenamefont
  {Dunham}}]{bartnikOperationalExperienceNanocoulomb2015}%
  \BibitemOpen
  \bibfield  {author} {\bibinfo {author} {\bibfnamefont {A.}~\bibnamefont
  {Bartnik}}, \bibinfo {author} {\bibfnamefont {C.}~\bibnamefont {Gulliford}},
  \bibinfo {author} {\bibfnamefont {I.}~\bibnamefont {Bazarov}}, \bibinfo
  {author} {\bibfnamefont {L.}~\bibnamefont {Cultera}},\ and\ \bibinfo {author}
  {\bibfnamefont {B.}~\bibnamefont {Dunham}},\ }\bibfield  {title} {\bibinfo
  {title} {Operational experience with nanocoulomb bunch charges in the
  {{Cornell}} photoinjector},\ }\href
  {https://doi.org/10.1103/PhysRevSTAB.18.083401} {\bibfield  {journal}
  {\bibinfo  {journal} {Phys. Rev. ST Accel. Beams}\ }\textbf {\bibinfo
  {volume} {18}},\ \bibinfo {pages} {083401} (\bibinfo {year}
  {2015})}\BibitemShut {NoStop}%
\bibitem [{\citenamefont {Gulliford}\ \emph {et~al.}(2013)\citenamefont
  {Gulliford}, \citenamefont {Bartnik}, \citenamefont {Bazarov}, \citenamefont
  {Cultrera}, \citenamefont {Dobbins}, \citenamefont {Dunham}, \citenamefont
  {Gonzalez}, \citenamefont {Karkare}, \citenamefont {Lee}, \citenamefont {Li},
  \citenamefont {Li}, \citenamefont {Liu}, \citenamefont {Maxson},
  \citenamefont {Nguyen}, \citenamefont {Smolenski},\ and\ \citenamefont
  {Zhao}}]{gullifordDemonstrationLowEmittance2013}%
  \BibitemOpen
  \bibfield  {author} {\bibinfo {author} {\bibfnamefont {C.}~\bibnamefont
  {Gulliford}}, \bibinfo {author} {\bibfnamefont {A.}~\bibnamefont {Bartnik}},
  \bibinfo {author} {\bibfnamefont {I.}~\bibnamefont {Bazarov}}, \bibinfo
  {author} {\bibfnamefont {L.}~\bibnamefont {Cultrera}}, \bibinfo {author}
  {\bibfnamefont {J.}~\bibnamefont {Dobbins}}, \bibinfo {author} {\bibfnamefont
  {B.}~\bibnamefont {Dunham}}, \bibinfo {author} {\bibfnamefont
  {F.}~\bibnamefont {Gonzalez}}, \bibinfo {author} {\bibfnamefont
  {S.}~\bibnamefont {Karkare}}, \bibinfo {author} {\bibfnamefont
  {H.}~\bibnamefont {Lee}}, \bibinfo {author} {\bibfnamefont {H.}~\bibnamefont
  {Li}}, \bibinfo {author} {\bibfnamefont {Y.}~\bibnamefont {Li}}, \bibinfo
  {author} {\bibfnamefont {X.}~\bibnamefont {Liu}}, \bibinfo {author}
  {\bibfnamefont {J.}~\bibnamefont {Maxson}}, \bibinfo {author} {\bibfnamefont
  {C.}~\bibnamefont {Nguyen}}, \bibinfo {author} {\bibfnamefont
  {K.}~\bibnamefont {Smolenski}},\ and\ \bibinfo {author} {\bibfnamefont
  {Z.}~\bibnamefont {Zhao}},\ }\bibfield  {title} {\bibinfo {title}
  {Demonstration of low emittance in the {{Cornell}} energy recovery linac
  injector prototype},\ }\href {https://doi.org/10.1103/PhysRevSTAB.16.073401}
  {\bibfield  {journal} {\bibinfo  {journal} {Phys. Rev. ST Accel. Beams}\
  }\textbf {\bibinfo {volume} {16}},\ \bibinfo {pages} {073401} (\bibinfo
  {year} {2013})}\BibitemShut {NoStop}%
\bibitem [{\citenamefont {Gulliford}\ \emph {et~al.}(2015)\citenamefont
  {Gulliford}, \citenamefont {Bartnik}, \citenamefont {Bazarov}, \citenamefont
  {Dunham},\ and\ \citenamefont
  {Cultrera}}]{gullifordDemonstrationCathodeEmittance2015}%
  \BibitemOpen
  \bibfield  {author} {\bibinfo {author} {\bibfnamefont {C.}~\bibnamefont
  {Gulliford}}, \bibinfo {author} {\bibfnamefont {A.}~\bibnamefont {Bartnik}},
  \bibinfo {author} {\bibfnamefont {I.}~\bibnamefont {Bazarov}}, \bibinfo
  {author} {\bibfnamefont {B.}~\bibnamefont {Dunham}},\ and\ \bibinfo {author}
  {\bibfnamefont {L.}~\bibnamefont {Cultrera}},\ }\bibfield  {title} {\bibinfo
  {title} {Demonstration of cathode emittance dominated high bunch charge beams
  in a {{DC}} gun-based photoinjector},\ }\href
  {https://doi.org/10.1063/1.4913678} {\bibfield  {journal} {\bibinfo
  {journal} {Appl. Phys. Lett.}\ }\textbf {\bibinfo {volume} {106}},\ \bibinfo
  {pages} {094101} (\bibinfo {year} {2015})}\BibitemShut {NoStop}%
\bibitem [{\citenamefont {Gulliford}\ \emph {et~al.}(2016)\citenamefont
  {Gulliford}, \citenamefont {Bartnik},\ and\ \citenamefont
  {Bazarov}}]{gullifordMultiobjectiveOptimizationsNovel2016}%
  \BibitemOpen
  \bibfield  {author} {\bibinfo {author} {\bibfnamefont {C.}~\bibnamefont
  {Gulliford}}, \bibinfo {author} {\bibfnamefont {A.}~\bibnamefont {Bartnik}},\
  and\ \bibinfo {author} {\bibfnamefont {I.}~\bibnamefont {Bazarov}},\
  }\bibfield  {title} {\bibinfo {title} {Multiobjective optimizations of a
  novel cryocooled dc gun based ultrafast electron diffraction beam line},\
  }\href {https://doi.org/10.1103/PhysRevAccelBeams.19.093402} {\bibfield
  {journal} {\bibinfo  {journal} {Physical Review Accelerators and Beams}\
  }\textbf {\bibinfo {volume} {19}},\ \bibinfo {pages} {093402} (\bibinfo
  {year} {2016})}\BibitemShut {NoStop}%
\bibitem [{\citenamefont {{van der Geer}}\ and\ \citenamefont {{de
  Loos}}(1997)}]{GPTSite}%
  \BibitemOpen
  \bibfield  {author} {\bibinfo {author} {\bibfnamefont {S.~B.}\ \bibnamefont
  {{van der Geer}}}\ and\ \bibinfo {author} {\bibfnamefont {M.~J.}\
  \bibnamefont {{de Loos}}},\ }\bibfield  {title} {\bibinfo {title}
  {Applications of the general particle tracer code},\ }in\ \href
  {https://doi.org/10.1109/PAC.1997.751279} {\emph {\bibinfo {booktitle}
  {Proceedings of the 1997 Particle Accelerator Conference (Cat.
  No.97CH36167)}}},\ Vol.~\bibinfo {volume} {2}\ (\bibinfo {year} {1997})\ pp.\
  \bibinfo {pages} {2577--2579 vol.2}\BibitemShut {NoStop}%
\bibitem [{\citenamefont {Maxson}\ \emph {et~al.}(2013)\citenamefont {Maxson},
  \citenamefont {Bazarov}, \citenamefont {Wan}, \citenamefont {Padmore},\ and\
  \citenamefont
  {{Coleman-Smith}}}]{maxsonFundamentalPhotoemissionBrightness2013}%
  \BibitemOpen
  \bibfield  {author} {\bibinfo {author} {\bibfnamefont {J.~M.}\ \bibnamefont
  {Maxson}}, \bibinfo {author} {\bibfnamefont {I.~V.}\ \bibnamefont {Bazarov}},
  \bibinfo {author} {\bibfnamefont {W.}~\bibnamefont {Wan}}, \bibinfo {author}
  {\bibfnamefont {H.~A.}\ \bibnamefont {Padmore}},\ and\ \bibinfo {author}
  {\bibfnamefont {C.~E.}\ \bibnamefont {{Coleman-Smith}}},\ }\bibfield  {title}
  {\bibinfo {title} {Fundamental photoemission brightness limit from disorder
  induced heating},\ }\href {https://doi.org/10.1088/1367-2630/15/10/103024}
  {\bibfield  {journal} {\bibinfo  {journal} {New J. Phys.}\ }\textbf {\bibinfo
  {volume} {15}},\ \bibinfo {pages} {103024} (\bibinfo {year}
  {2013})}\BibitemShut {NoStop}%
\bibitem [{\citenamefont {Pierce}\ \emph {et~al.}(2020)\citenamefont {Pierce},
  \citenamefont {Andorf}, \citenamefont {Lu}, \citenamefont {Gulliford},
  \citenamefont {Bazarov}, \citenamefont {Maxson}, \citenamefont {Gordon},
  \citenamefont {Kim}, \citenamefont {Norvell}, \citenamefont {Dunham},\ and\
  \citenamefont {Raubenheimer}}]{pierceLowIntrinsicEmittance2020}%
  \BibitemOpen
  \bibfield  {author} {\bibinfo {author} {\bibfnamefont {C.~M.}\ \bibnamefont
  {Pierce}}, \bibinfo {author} {\bibfnamefont {M.~B.}\ \bibnamefont {Andorf}},
  \bibinfo {author} {\bibfnamefont {E.}~\bibnamefont {Lu}}, \bibinfo {author}
  {\bibfnamefont {C.}~\bibnamefont {Gulliford}}, \bibinfo {author}
  {\bibfnamefont {I.~V.}\ \bibnamefont {Bazarov}}, \bibinfo {author}
  {\bibfnamefont {J.~M.}\ \bibnamefont {Maxson}}, \bibinfo {author}
  {\bibfnamefont {M.}~\bibnamefont {Gordon}}, \bibinfo {author} {\bibfnamefont
  {Y.-K.}\ \bibnamefont {Kim}}, \bibinfo {author} {\bibfnamefont {N.~P.}\
  \bibnamefont {Norvell}}, \bibinfo {author} {\bibfnamefont {B.~M.}\
  \bibnamefont {Dunham}},\ and\ \bibinfo {author} {\bibfnamefont {T.~O.}\
  \bibnamefont {Raubenheimer}},\ }\bibfield  {title} {\bibinfo {title} {Low
  intrinsic emittance in modern photoinjector brightness},\ }\bibfield
  {journal} {\bibinfo  {journal} {Phys. Rev. Accel. Beams}\ }\textbf {\bibinfo
  {volume} {23}},\ \href {https://doi.org/10.1103/physrevaccelbeams.23.070101}
  {10.1103/physrevaccelbeams.23.070101} (\bibinfo {year} {2020})\BibitemShut
  {NoStop}%
\bibitem [{\citenamefont {Kim}(1989)}]{kimRfSpacechargeEffects1989}%
  \BibitemOpen
  \bibfield  {author} {\bibinfo {author} {\bibfnamefont {K.-J.}\ \bibnamefont
  {Kim}},\ }\bibfield  {title} {\bibinfo {title} {Rf and space-charge effects
  in laser-driven rf electron guns},\ }\href
  {https://doi.org/10.1016/0168-9002(89)90688-8} {\bibfield  {journal}
  {\bibinfo  {journal} {Nuclear Instruments and Methods in Physics Research
  Section A: Accelerators, Spectrometers, Detectors and Associated Equipment}\
  }\textbf {\bibinfo {volume} {275}},\ \bibinfo {pages} {201} (\bibinfo {year}
  {1989})}\BibitemShut {NoStop}%
\end{thebibliography}%

\end{document}